\documentclass[a4paper]{article}
\usepackage[table]{xcolor}
\usepackage{algorithm, algpseudocode}
\usepackage{amsmath, amssymb}
\usepackage{authblk}
\usepackage{booktabs}
\usepackage{physics}
\usepackage[labelfont=bf, font=small]{caption}
\usepackage{enumitem}
\usepackage{graphicx}
\usepackage{hyperref}
\usepackage{ragged2e} 
\usepackage[counterclockwise]{rotating}
\usepackage{svg}
\usepackage{url}
\usepackage{comment}
\usepackage{cleveref} % must be after hyperref
\usepackage[top=2cm, bottom=2cm, right=2cm, left=2cm]{geometry}

\usepackage[font=footnotesize, labelfont=bf, justification=justified]{caption}
\usepackage{subcaption}
\hypersetup{
    colorlinks=true,
    linkcolor=blue,
    filecolor=magenta,      
    urlcolor=cyan,
    citecolor=blue
}

\title{\textbf{Supervised Similarity for High-Yield Corporate Bonds with Quantum Cognition Machine Learning}}

\author[1, *]{Joshua Rosaler}
\author[2, 3]{Luca Candelori}
\author[2]{Vahagn Kirakosyan}
\author[2]{Kharen Musaelian}
\author[2]{Ryan Samson}
\author[4]{Martin T. Wells}
\author[1]{Dhagash Mehta}
\author[1]{Stefano Pasquali}

\affil[1]{BlackRock, Inc., New York, NY}
\affil[2]{Qognitive, Inc., Miami Beach, FL}
\affil[3]{Wayne State University, Department of Mathematics, Detroit, MI}
\affil[4]{Cornell University, Department of Statistics and Data Science, Ithaca, NY}

\affil[*]{Primary Investigator (\href{mailto:Joshua.Rosaler@blackrock.com}{Joshua.Rosaler@blackrock.com})}
\date{\today}

\begin{document}
\maketitle

\begin{abstract}
We investigate the application of quantum cognition machine learning (QCML), a novel paradigm for both supervised and unsupervised learning tasks rooted in the mathematical formalism of quantum theory, to distance metric learning in corporate bond markets. Compared to equities, corporate bonds are relatively illiquid and both trade and quote data in these securities are relatively sparse. Thus, a measure of distance/similarity among corporate bonds is particularly useful for a variety of practical applications in the trading of illiquid bonds, including the identification of similar tradable alternatives, pricing securities with relatively few recent quotes or trades, and explaining the predictions and performance of ML models based on their training data. Previous research has explored supervised similarity learning based on classical tree-based models in this context; here, we explore the application of the QCML paradigm for supervised distance metric learning in the same context, showing that it outperforms classical tree-based models in high-yield (HY) markets, while giving comparable or better performance (depending on the evaluation metric) in investment grade (IG) markets.
\end{abstract}

\section{Introduction}

The learning of similarity or distance relations between securities is a central area in the application of machine learning to financial contexts. For example, a portfolio manager may wish to include a particular security in a portfolio but then find that there is insufficient liquidity in the market, and instead try to find a substitute that is as similar as possible to the desired security that can be traded. Alternatively, one can also use similarity to price securities for which there are no recent trades or quotes in the market by looking at the prices of similar securities with more up-to-date market data. In addition, one can use a measure of similarity between securities for purposes of hedging risk, for example by going long one security and shorting securities that are similar to it. Yet another application is anomaly detection, where one may wish to identify securities within a particular cohort that are not typical of the cohort as those that are most dissimilar from the others. Beyond the individual security level, one can also use distance metric learning to assess similarity of different portfolios; for example in basket trading one may wish to assess the similarity of an entire basket of securities to a given ETF. The list of potential applications goes on. 
%Particularly in less liquid markets, where data is sparse and securities can be harder to trade, a meaningful measure of similarity can be essential to identifying tradable alternatives for securities with no available counterparty, to determining a fair price for such securities in cases where there are few recent trades or quotes, to explaining the predictions of ML models based on the most similar training points, identification of outlier securities within financial categorization systems such as GICS, and many others. 

While most work in the field of similarity or distance metric learning falls under the banner of unsupervised learning, for financial contexts such as those mentioned above, a supervised approach to similarity, in which the notion of similarity is defined specifically with respect to the target variable of a supervised ML model, can be more effective. For example, a portfolio manager may choose to include a particular bond in a portfolio based on its spread, or excess yield relative to treasury yield associated with the same maturity. If that security is hard to trade, the PM may settle instead for an alternative security with comparable spread that also has a similar risk profile based on fundamental features such as coupon, rating, sector, time to maturity, etc. Recent work has shown that by training tree-based ensemble models such as a random forest or gradient boosted machine (GBM) to predict a given target variable, one can extract a distance metric between points in the feature space such that nearest neighbors under this metric also have similar values for the target variable \cite{breiman-cutler-blog}, \cite{hastie2009random}, \cite{liaw2002classification}, \cite{rhodes2023geometry}. Financial applications of supervised similarity based on random forests and GBMs have recently been explored in \cite{desai2023quantifying},\cite{jeyapaulraj2022supervised},\cite{rosaler2024enhanced},\cite{feng2024open},\cite{castellanos2024can},\cite{li2024quantile}, \cite{yampolsky2024case}, \cite{saha2024machine}. Currently, supervised similarity based on random forests represent the state-of-the-art for supervised similarity, as there exists a substantial literature on this method together with multiple libraries and methods for computing RF similarities \cite{rfgap}, \cite{rfproximity}.

One can evaluate the quality of such a metric by evaluating the performance of a $k$-nearest-neighbors regressor 
\footnote{Similar remarks also apply in the context of classification models.},
where the notion of ``nearest" is taken from the learned metric. The intuition behind this evaluation strategy is that, in cases where we want our notion of distance on the feature space to reflect closeness in the values of a particular target variable, a metric is better if points that are closer under the metric tend to have values for the target variable that are also closer to that of the test point. In such cases, an average of training targets for nearest neighbors in the training set will tend to have lower error in predicting the value of the prediction target for the test point than the same average for a distance metric that was not tailored specifically to this target.  

The emerging field of quantum cognition machine learning (QCML) \cite{MusaelianEtAl, CandeloriEtAl, SamsonEtAl} offers a new class of machine learning models that is rooted in the mathematical formalism of quantum theory, in which individual data points are represented by quantum states in a complex Hilbert space, and features and target variables are represented by Hermitian operators or ``observables" that are learned by optimizing a particular objective function over the observables' parameters. While the objective functions and evaluation metrics of QCML models are largely the same as those used in the training and evaluation of classical machine learning models, the way in which data are represented and in which models of the functional dependencies among features are parametrized, deviates in fundamental ways from classical machine learning methods. In particular, in this article we introduce a new notion of proximity between data points that arises naturally in QCML by taking the quantum fidelity - i.e., the absolute value of the inner product - of two quantum states, each representing a data point. 

A QCML model, just like a random forest, can naturally handle both numerical and categorical data, as well as missing and/or noisy data. It creates a global quantum manifold model (in the sense of quantum geometry \cite{Ishiki_2015, Steinacker_2021, Steinacker_book}) of the original data manifold, that is robust to noise and able to generalize well beyond training samples \cite{CandeloriEtAl, SamsonEtAl}. Part of this adeptness at controlling variance stems from the fact that the number of parameters of a QCML model scales only quadratically with the number of features, thus achieving logarithmic economy of representation. By way of comparison, the depth of trees in a random forest typically scales linearly with the number of features, resulting in an exponential growth of parameters/leafs. These different scaling regimes result in profound differences between random forest proximity and QCML proximity. For example, we show in this article by way of example that in the presence of a large number of features and a large number of outliers in the data, random forest proximity tends to inflate sparsity in the data by placing points as far apart from each other as possible. This is a consequence of the random forest model attempting to grow individual tree branches for each outlier, thus over-parametrizing the data. On the other hand, QCML proximity creates a compact representation of the data even in the presence of a large number of outliers. In this article, these effects are shown to be particularly dramatic when analyzing a cohort of high-yield bonds, which exhibit a) a large number of features arising from one-hot encoding of categorical variables (bond rating, country, and industry) and b) a large number of outliers, arising from bonds that are near default. In this setting, we demonstrate a significant advantage of QCML proximity over random forest proximity, as measured by a $k$-nearest neighbors regressor. For control, we also analyze a data set consisting of investment-grade bonds, and we show that the performance of QCML and random forest is similar. This is as expected, since investment grade bonds tend to have a more compact distribution and a smaller number of outliers compared to high-yield bonds.

% At a high level, the advantage conferred by this alternative paradigm comes from the non-commutativity that it allows between different features. \textcolor{red}{@Luca Can we elaborate on this to say a bit more intuitively what the source of QCML's advantages are, and in which specific sorts of context we think it is likely to outshine classical ML? For example, with deep neural nets one often hears that they shine in the modeling of unstructured data, high dimensions, and that this is in part to do with the fact that successively deeper layers learn successively higher-level, more coarse-grained abstractions in the data. Is there an analogous story one can tell about QCML?}

% In this article, we demonstrate the effectiveness of QCML in the context of distance metric learning for corporate bonds, showing that it gives performance that is better than or comparable to start-of-the-art supervised similarity methods based on random forests. Section \ref{section:supervised_metric} introduces supervised distance metric learning with random forests.
The article is organized as follows. Section \ref{section:supervised_metric} reviews previous work on the application of classical tree-based methods of supervised similarity. Section \ref{section:qcml} introduces the basics of QCML, particularly as applied to the context of regression and supervised similarity learning. Section \ref{section:data} describes the corporate bond dataset on which we train our models. Section \ref{section:methods} describes the detailed methodology around training and evaluation of our models. Section \ref{section:results} presents results that compare the performance of both the classical and QCML metrics. Section \ref{section:conclusion}, the conclusion, summarizes our findings and describes potential areas for further exploration.

\section{Supervised Metric Learning with Random Forests} 
\label{section:supervised_metric}

In the context of tree-based regression models such as random forests or gradient boosted decision trees, there are several ways to quantify the similarity of points of the feature space, all of which rely on the proportion of trees in the ensemble for which two points fall in the same leaf node. Here we focus on the case of the random forest, which is much more developed, both in terms of research literature and available libraries, than for GBMs. 

\subsection{Breiman's Original Definition}

Breiman originally proposed a definition of supervised similarity for random forests as the proportion of trees in the RF ensemble for which two points fall in the same leaf node:

\begin{equation}\label{eq:breiman_prox}
Prox(i,j) = \frac{1}{M} \sum_{T=1}^{M} \frac{1}{N_{i}^{T}}I[j \in \mathcal{L}_{i}^{T}]
\end{equation}

\noindent where $I$ is the indicator function, $M$ is the number of trees in the forest and $N_{i}^{T}$ is the number of training points in the leaf $\mathcal{L}_{i}^{T}$ of tree $T$ into which point $i$ and $j$ both fall \cite{breiman-cutler-blog}.

\subsection{Out-of-Bag Proximities}
Breiman's original definition of RF-based proximity does not differentiate between in-bag and out-of-bag data, which may artificially inflate the similarity between points in the same bootstrap sample. To address this shortcoming, a revised notion of RF-based similarity, the out-of-bag (OOB) proximity, was introduced \cite{hastie2009random,liaw2002classification}. This notion of proximity refines Breiman's definition by considering only pairs of points that are out-of-bag for each tree:

\begin{equation}\label{eq:oob}
Prox_{OOB}(i,j) = \frac{\sum_{t\in{S_{i}}}I(j\in{O(t)\cap{\upsilon(t)}})}{\sum_{t\in{S_{i}}}I(j\in{O(t)})},
\end{equation}

\noindent where $O(t)$ denotes the set of data indices that are out-of-bag samples in the $t^{th}$ tree, $S_{i}$ is the set of trees where the $i^{th}$ observation is out-of-bag, and $\upsilon(t)$ denotes the indices of data records that end up in the same terminal leaf node as $x_{i}$ in the $t^{th}$ tree.

\subsection{Geometry-and-Accuracy-Preserving (GAP) Proximities}

Ref.\cite{lin2006random} showed that any tree-based ensemble model, such as a random forest or gradient boosted machine, can be viewed as an adaptive weighted $k$-nearest-neighbors (KNN) model. That is, the prediction of the RF or GBM for any test point can be expressed exactly as a weighted average of the target labels of points in the training set, where the weights of the training points vary depending on the location of the test point in the feature space. In the context of regression, this result entails that one can expand the prediction $\hat{y}_{i}$ of the model locally as a linear combination of target labels in the training dataset:

\begin{equation}\label{eq:RF_weighted_KNN}
\hat{y}_{i} = \textbf{k}_{i}(x) \cdot \textbf{y}_{train} = k_{i,1}(x) \ y_{train,1} \ + \ ... \ + \ k_{i,N}(x) \ y_{train,N},
\end{equation}
where $y_{train, j}$ is the ground truth target label for the $j^{th}$ training example, and $k_{i,j}(x)$ is the input-dependent weight of the observation $j$ in the expansion for observation $i$. This weight corresponds to yet another notion of proximity or similarity between points in the feature space, and, like the other proximities we have discussed, depends on the number of trees in the RF ensemble for which two points fall in the same leaf node \cite{rhodes2023geometry}.

Ref.~\cite{rhodes2023geometry} has shown that for random forests, the correct form of the expansion coefficients $k_{ij}(x)$ appearing in Eq (\ref{eq:RF_weighted_KNN}), which the authors
call the Geometry and Accuracy Preserving (GAP) RF proximity,  is

\begin{equation} \label{GAP}
Prox_{GAP}(i,j) = \frac{1}{|S_{i}|} \sum_{t \in S_{i}} \frac{c_j(t)I[j \in J_{i}(t)]}{|M_{i}(t)|},
\end{equation}
where $S_{i}$ is the set of trees in the RF for which observation $i$ is out of bag, $M_i(t)$ is the multiset 
\footnote{Recall that a multiset is a generalization of the concept of a set, allowing for repetition among the elements of the set, where the number of repetitions of a unique element in the multiset is known as its multiplicity.}
of bagged points in the same leaf as $i$ in tree $t$, $J_{i}(t)$ is the corresponding set (i.e., without repetitions) of bagged points in the same leaf as $i$ in tree $t$, and $c_{j}(t)$ is the multiplicity of the index $j$ in the bootstrap sample.

\section{Quantum Cognition Machine Learning}
 \label{section:qcml}

 Quantum Cognition Machine Learning (QCML) has recently been proposed as a new framework for statistical learning models based on ideas from quantum cognition \cite{MusaelianEtAl, CandeloriEtAl, SamsonEtAl}. The origins of quantum cognition can be traced back to the works of Aerts et al \cite{Aerts1995-AERAOQ}, Khrennikov \cite{KHRENNIKOV2006225}, and Busemayer et al \cite{Busemeyer_Bruza_2012} (see \cite{Porthos_Busemayer_2022} for a recent survey). 
 %In these works, it is posited that the state of mind is formally given by a quantum state, and all questions that can be answered within that state of mind are represented as quantum observables. 
 
Similarly, QCML models learn a representation of the data into quantum states.
%, and make forecasts by taking quantum measurements. 
To illustrate the learning and forecasting process in detail, recall that in quantum mechanics a state is a vector of unit norm in a Hilbert space, and is represented in bra-ket notation by a ket $\ket{\psi}$. The inner product of two states $\ket{\psi_1}$, $\ket{\psi_2}$ is represented by a bra-ket $\braket{\psi_1}{\psi_2}$. The expectation value of a Hermitian operator $M$ (i.e. a quantum observable) on a state $\ket{\psi}$ is denoted by $\expval{M}{\psi}$, representing the expected outcome of the measurement corresponding to $M$ on the state $\ket{\psi}$. In this interpretation, the eigenvalues of $M$ represent the possible outcomes of the measurement, while the probabilities of each outcome occurring are given by the norm squared of the braket between $\psi$ and the eigenvector corresponding to the outcome. Therefore, the expression $\expval{M}{\psi}$ corresponds exactly to the expected value of the discrete random variable given by measuring $M$ on $\psi$ \cite[I.2.2]{nielsen00}. 

In QCML, for each vector $\mathbf{x}_t \in \mathbb{R}^K$ belonging to a data set consisting of $t=1, \ldots, T$ $K$-dimensional observations, define an error Hamiltonian 
\begin{equation}
    \label{eq:square_hamiltonian}
    H(\mathbf{x}_{t}) = \frac{1}{2}\sum_k (A_k - \mathbf{x}_{t,k} \cdot I)^2
\end{equation}
depending on a fixed set of $N$-dimensional quantum observables $\{A_k\}$. In this formula, $I$ denotes the $N\times N$ identity matrix. Each of these $K$ quantum observables can be viewed as a `quantization' of a corresponding feature of the original $K$-dimensional data set. The vector $\mathbf{x}_t$ is then mapped to the ground state $\ket{\psi_t}$ of the error Hamiltonian (i.e. the eigenstate associated with the lowest eigenvalue), giving 
%\$ l.543 Linear Regression \& \textbackslash{}infty \& .57 \{\scriptsize \$\pm\$ .03\} \&... I've inserted a begin-math/end-math symbol since I think you left one out. Proceed, with fingers crossed.
%\subsubsection{Missing \$ inserted.}
a representation of data into quantum states (i.e., normalized vectors in a complex Hilbert space). Conversely, given an arbitrary $N$-dimensional quantum state $\ket{\psi}$, we can define its `position' to be the $K$-dimensional vector 
$$
\mathbf{x}(\psi) =   (\expval{A_k}{\psi})_{k} \in \mathbb{R}^K, 
$$
which in quantum theory represents the expected outcome of measuring the quantum observables $A_k$ on $\psi$. In this way, given a set of quantum observables $\{A_k\}$, we have a way to map data into quantum states by sending $\mathbf{x}_t$ to its ground state $\ket{\psi_t}$, and we can also retrieve information about a quantum state $\ket{\psi}$ by taking its position $\mathbf{x}(\psi)$. In an unsupervised setting, training a QCML model involves iterative updates to the observables $\{A_k\}$ so that the ground states $\ket{\psi_t}$ `cohere' to the data, that is, the distance between $\mathbf{x}_t$ and its position $\mathbf{x}(\psi_t)$ is minimized, as well as the variance of the measurement. This can be achieved by minimizing this distance directly, or by minimizing the overall energy of the error Hamiltonian, as discussed in detail in  \cite{CandeloriEtAl}. 

In the supervised setting, which is the primary focus of this article, the training process is slightly different, and is discussed in detail in \cite{SamsonEtAl}. The target variable $y \in \mathbb{R}$ is assigned a $N$-dimensional quantum `forecast' observable $B$, and given a data point $\mathbf{x}_t$ the corresponding forecast is given by
$$
\widehat{y}_t = \expval{B}{\psi_t}.
$$
During the training process, the quantum observables $\{A_k\}$ and $B$ are updated at each iteration so as to minimize the mean absolute error $\sum_t |\widehat{y}_t - y_t|$. Other loss functions are possible, such as mean squared error, but in practice MAE seems to lead to more accurate forecasts. The non-differentiability of the loss function here does not add further complexity to the algorithm, since the mapping from $\mathbf{x}_t$ to its ground state $\psi_t$ is already non-differentiable, the singular points corresponding to the locus of degeneracy of the error Hamiltonian \eqref{eq:square_hamiltonian}. This setup can be easily extended to the case of multiple target variables. 

The training algorithm can be summarized as follows:

\begin{algorithm}
    \caption*{\bf{QCML univariate regression model training}}
    \label{alg:training}
    \begin{algorithmic}
    \State \textbullet\ Randomly initialize feature operators $\{A_k\}$ and target operator $B$.
    \State \textbullet\ Iterate over training data and operators until desired convergence:
    \begin{algorithmic} \item
        \begin{algorithmic}[1]
        \State Generate error Hamiltonian $H(\mathbf{x}_{t})$
        \State Holding $A_k$ constant, find the ground state $\ket{\psi_t}$ of $H(\mathbf{x}_{t})$
        \State Calculate gradients of the loss function $\sum_t |\widehat{y}_t - y_t|$ w.r.t $A_k$ and $B$
        \State Update $A_k$ and $B$ via gradient descent
        \end{algorithmic}
    \end{algorithmic}
    \end{algorithmic}
\end{algorithm}

The specifics of each of these steps will depend on the choice of parametrization for the operators $A_k$ and $B$, and different choices of loss function and optimization process are possible. The dimension of the Hilbert space $N$ is a hyper-parameter of the algorithm, and can be optimized using cross-validation. Larger values of $N$ typically lower the loss, but could lead to overfitting and worse out-of-sample performance, while lower dimensions tend to have higher bias and lower variance \cite{CandeloriEtAl}. In practice, it is also best to keep $N$ small for computational efficiency.

We should point out that the Hamiltonian \eqref{eq:square_hamiltonian} and its ground states also appear prominently in the field of theoretical physics known as Quantum Geometry  \cite{Ishiki_2015, Steinacker_2021, Steinacker_book}. In this setting, the set of observables $\{A_k\}$ is called a `matrix configuration', and is typically fixed, given by well-known quantum operators (e.g. the spin operators quantizing angular momentum). The main goal in this context is to construct a Poisson manifold that is quantized by the matrix configuration. This is somewhat opposite to the goal of QCML, which instead learns the matrix configuration from the data manifold. Nevertheless, the field of Quantum Geometry offers an interesting point of intersection between theoretical quantum physics and QCML. 

\subsection{QCML proximity}
\label{sec:QCML_distance}

There is a natural notion of proximity for quantum states given by {\em quantum fidelity} \cite[III.9]{nielsen00}
$$
f(\psi_1, \psi_2) = \lvert \braket{\psi_1}{\psi_2} \rvert ^ 2,
$$
which can be interpreted as the probability of identifying the state $\psi_1$ with the state $\psi_2$, when performing a quantum measurement designed to test whether a given quantum state is equal to $\psi_2$ (or vice-versa). 

In the context of QCML, this type of proximity can be used to define a similarity measure on the data. Indeed, suppose that a (supervised) QCML model with feature observables $A_k$ and target observable $B$ has been trained.  We then have a representation of the data in quantum states given by $\mathbf{x}_t \rightarrow \ket{\psi_t}$. Given two data points $\mathbf{x}_t, \mathbf{x}_{t'}$, define their {\em QCML distance} by 
\begin{equation}
\label{eqn:QCML_distance}
d_Q(\mathbf{x}_t, \mathbf{x}_{t'}) = 1 - f(\psi_t, \psi_{t'}) = 1 - \lvert \braket{\psi_t}{\psi_{t'}} \rvert ^ 2.
\end{equation}
Note that in contrast to the standard Euclidean distance between two data points, the QCML distance is a type of supervised similarity measure, since the data representation into quantum states $\psi_t$ has been optimized using the training targets. In particular, just like with Random Forest proximity, it is possible to apply QCML proximity to data containing both numerical and categorical features. 

% %\subsection{Feature importance for QCML similarity}
% %
% The simple analytic formula \eqref{eqn:QCML_distance} for QCML distance lends itself to interpretability, in terms of the gradients of quantum fidelity with respect to each data feature. To calculate these gradients, suppose we have two data vectors $\mathbf{x}_t, \mathbf{x}_{t'}$ and we want to know how much a small change in the feature values of $\mathbf{x}_t$ affects the QCML distance $ d_Q(\mathbf{x}_t, \mathbf{x}_{t'})$. For simplicity, let $\psi = \psi_{t}$ and $\phi = \psi_{t'}$. Then the first-order effect on QCML distance of variation of the $k$-th feature value of $\mathbf{x}_t$ can be calculated exactly by the exact formula
% \begin{align*}
% \pdv{d_Q}{x_k} &= \pdv{d_Q}{\psi}\pdv{\psi}{x_k} + \pdv{d_Q}{\psi^*}\pdv{\psi^*}{x_k} \\
% &= -2 \Re \Bigg( \braket{\psi}{\phi}\sum_{n \neq 0}\frac{\braket{\phi}{n}\bra{n}A_{k}\ket{\psi}}{E_n - E_0}\Bigg),
% \end{align*}
% where $E_n$ (resp. $\ket{n}$), $n=0,\ldots, N-1$ denote the eigenvalues (resp. eigenstates) of the error Hamiltonian $H(\mathbf{x}_t)$. This exact formula can easily be derived using first-order perturbation theory and Hellmann-Feynman-type formulas.  

\section{Data Description}
\label{section:data}

%\textcolor{red}{@Luca In Figure 1, maybe we should just show the distributions of the target variables price and spread for HYG and IGSB in one plot, and the distributions of any input features in another? Otherwise it seems confusing to present the targets alongside the input features in an undifferentiated way}  

We selected two data sets of corporate bonds consisting of the holdings of 1) the iShares IBoxx \$ High Yield ETF (ticker: HYG) and 2) iShares 1-5 Year Investment Grade ETF (ticker: IGSB). The holdings data of both ETFs are publicly available. The data features selected are cross-sectional, and they were downloaded at the end of day on 09/18/2024. They consist of seven numerical and three categorical variables, with the target variable being the bond yield for HYG and the credit spread for IGSB. A detailed list of features is given in Table \ref{table:list_of_features}. We use one-hot-econding for the categorical variables, obtaining a total of $D=99$ features. 

\begin{table}[H]
\centering

\begin{subtable}[b]{0.49\textwidth}
    \centering
\begin{tabular}{c|c}
        \bf{Data feature} & \bf{Type}\\
        \hline
        Coupon & numerical\\
        Coupon Frequency & numerical\\
        Days To Maturity & numerical\\
        Duration & numerical\\
        Age & numerical\\
        Amount Issued & numerical\\
        Amount Outstanding &numerical \\
        Rating & categorical\\
        Country & categorical\\
        Industry & categorical
    \end{tabular}
\end{subtable}
\begin{subtable}[b]{0.49\textwidth}
    \centering
    \begin{tabular}{c|c}
        \bf{Target variables } & \bf{Type}\\
        \hline
        Yield (for HYG) & numerical\\
        Credit Spread (for IGSB) & numerical\\
    \end{tabular}
\end{subtable}
\caption{List of features and target variables for corporate bond data.}
   \label{table:list_of_features}
\end{table}

We choose credit spread, the difference between the bond's yield and the yield of a treasury bond of comparable maturity, as the prediction target for IGSB because investment-grade bonds are typically traded on spread rather than yield. The spread reflects the additional yield that one receives in compensation for assuming the additional risk of default associated with a corporate bond as opposed to a treasury bond, which is considered ``risk-free." We choose the bond yield (return on initial investment up to time of maturity) rather than spread as the prediction target for HYG bonds as the higher level of risk and volatility associated with high-yield bonds makes it less useful to compare the yield of the HY bond to that of a treasury bond; it is more practical simply to consider the yield directly for these bonds. 

\begin{figure}[H]
\centering
\includegraphics[scale=0.4]{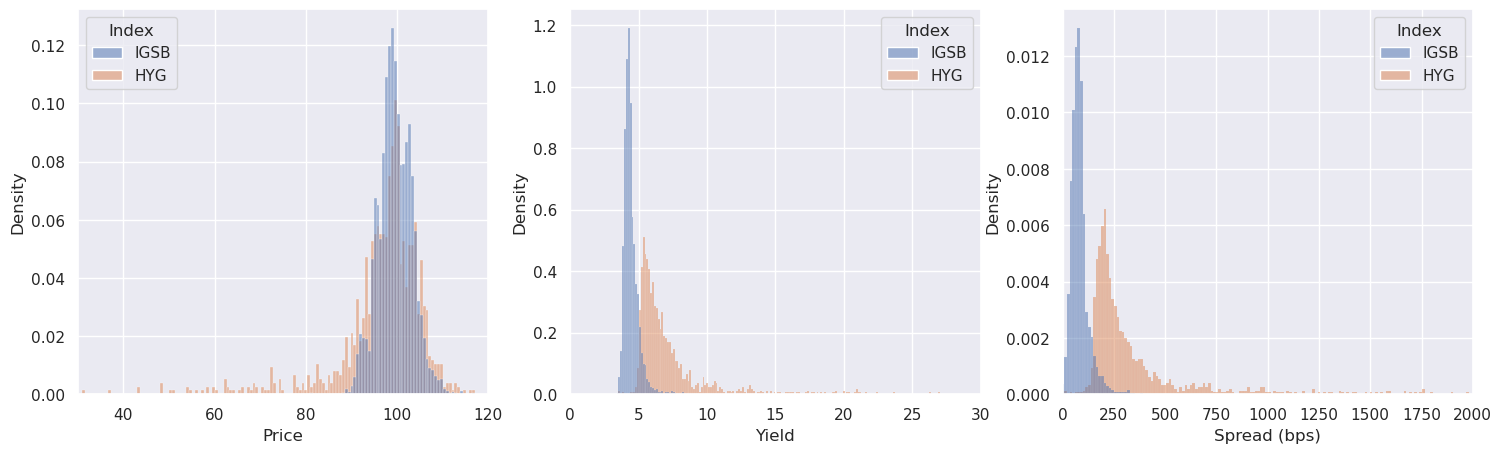}
\caption{Distributions of bond price, yield and spread for HYG and IGSB}
\label{fig:target_distribution}
\end{figure}

\begin{figure}[H]
\centering
\includegraphics[scale=0.4]{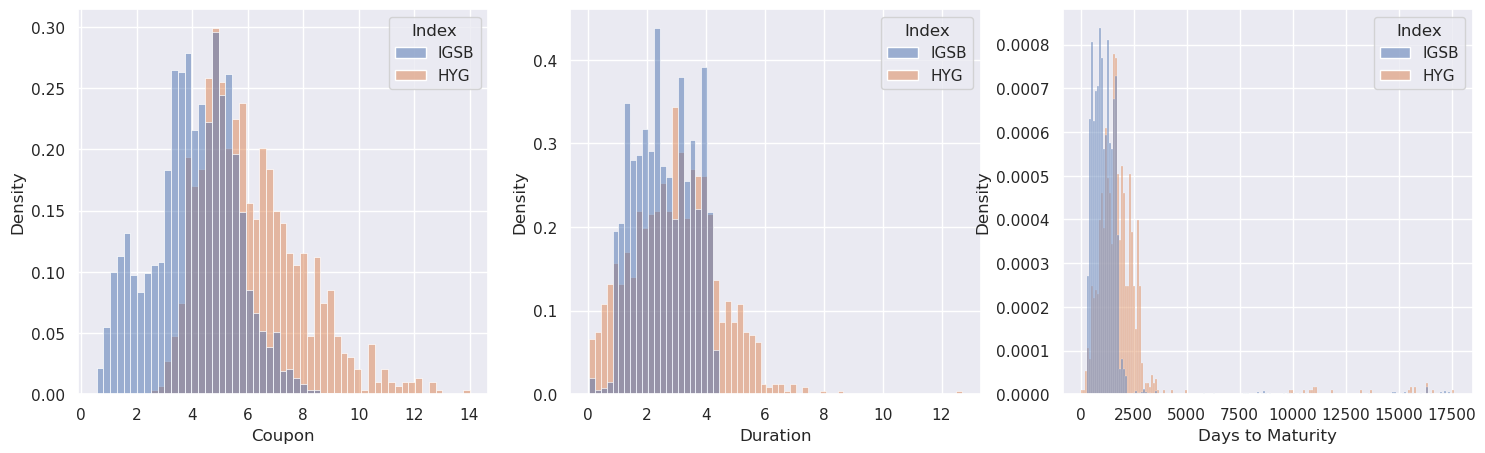}
\caption{Distribution of some of the data features for both HYG and IGSB}
\label{fig:data_distribution}
\end{figure}

There are significant differences in the distribution of the data between HYG and IGSB data. In general, the distributions of bond price, yield (HYG target label), and spread (IGSB target label) tend to have wider support and a relatively larger number of outliers for the high-yield bonds in HYG, compared to the investment grade bonds in IGSB \ref{fig:target_distribution}. Likewise, the distribution of features for high-yield bonds has wider support and tends to have a relatively larger number of outliers, compared to investment grade bonds(Figure \ref{fig:data_distribution}).

We will demonstrate that these substantial differences in the distributions of the data features lead to significant differences in predictive performance when comparing QCML and Random Forest. In particular, we will show below that QCML holds an advantage on HYG, due to the sparse characteristics of the data.

\section{Methods}
\label{section:methods}

We now describe how to train and evaluate the QCML model on the regression tasks  described above (i.e. yield prediction for HYG and spread prediction for IGSB), and how to calculate and evaluate the supervised QCML distance between data points. 

\subsection{Train-test split and hyperparameter optimization}

We first performed a random 80-20 train-test split of the data, and used the train data for hyperparameter optimization. For the QCML model, we optimized Hilbert space dimension by using a 3-fold cross-validation procedure, minimizing mean squared error (MSE). We obtained an optimal Hilbert space dimension of $N=7$ for HYG and $N=12$ for IGSB. 
%For the regression task, we combined an ensemble of 3 QCML models with different weight initializations, and took the average of the predictions.  

For comparison with Random Forest, we also did a hyperparameter search varying the number of trees, maximum depth, minimum samples per leaf, maximum number features considered in each split, and training objective function. The optimal choice of parameters for HYG was max. depth = 50, no. trees = 1000, minimum samples per leaf 1, number of features considered in each split equal to the square root of the total number of features, and training objective equal to mean absolute error, while for IGSB was max. depth = 50, no. trees = 200, minimum samples per leaf 1, number of features considered in each split equal to the square root of the total number of features, and training objective equal to mean squared error. 

\subsection{Similarity Matrix Evaluation}

In the context of unsupervised distance metric learning, it can be difficult to assess quantitatively whether one metric is ``better" than another as there is no prediction target. In the context of supervised distance metric learning, however, the existence of a prediction target enables more quantitative evaluation and comparison of the two distance metrics. To evaluate the quality of the similarity measure, we run the K-Nearest-Neighbors (KNN) regression method on the same train-test split using both the QCML distance and the standard Euclidean distance. The idea is that a good measure of distance should bring bonds of similar yield or spread closer and it should separate bonds with differing yield or spread, translating into better performance under KNN.  
%We use Python’s Sklearn package implementation of KNN. The implementation allows users to supply a custom distance metric, by supplying a pre-computed distance matrix between all the training points during fitting, and a distance matrix between train and test set for prediction. The QCML distance matrices can be calculated using formula \eqref{eqn:QCML_distance}. 
For the QCML metric, we ensembled 3 different distance matrices, corresponding to QCML models with different initialization weights, by taking the average of each entry, which is computed as the inner product $\lvert \braket{\psi_t}{\psi_{t'}} \rvert$. For the RF-based metrics, the proximities are given directly by the formulas \eqref{eq:breiman_prox}, \eqref{eq:oob}, \eqref{eq:RF_weighted_KNN}. For the Euclidean metric, the proximity is defined as $1-d$, where $d$ is the Euclidean distance normalized by its maximum training value. 
%With this implementation we used proximity-weighted KNN as compared to uniformly weighted, in order to emphasize the contribution of the custom metric towards the overall prediction. 

%by assessing the performance of the k-nearest neighbors prediction as a function of k, where ``nearest" is determined by the learned distance metric. Taking the case of HYG bonds as an example, the intuition behind this evaluation method is that, given we want the nearest neighbors of a given bond to be bonds of similar yield, a metric is better if the nearest neighbors to a given bond, according to that metric, have yields that are closer to that of the target bond on average. Thus, we assess the relative effectiveness of two metrics by plotting the out-of-sample performance of a knn regressor based on that metric as a function of k; the metric with the lower-lying curve is deemed to have superior performance. 

However, one subtlety remains as to whether and if so how one should weight the contributions of the $k$ nearest neighbors when computing the KNN prediction. In addressing this issue, it is essential to keep in mind our use case, which is to identify trade-able substitutes for a bond that one wishes to buy but can't. Ideally, such a bond would have not only similar yield, but also similar values for input features like credit rating, amount issued, industry, and so on, since these affect the various types of risk associated with holding the bond. It is of greater value to a trader or portfolio manager when a bond of similar yield to the desired bond also has similar values for these inputs features than when a potential substitute bond has similar yield but substantially different values for these input features. So, when averaging deviations between the yield of the target bond and those of its neighbors to assess overall performance of the metric, we should place greater weight on bonds with higher proximity, which will have more similar values for the most important input features. 

The problem remains as to how precisely one constructs such weights, which quantify the importance of a bond having values for the input features that are close to those of the target bond. Apart from the general assumption that they should generally fall off with distance $d$ (or equivalently, increase with proximity), the question of which weights one should use is largely subjective and ill-defined. For the Euclidean metric, does it make more sense to use $1-d$ (assuming distances have been normalized to be between 0 and 1), $1/d$, $1/d^{2}$, $e^{-d}$, etc. as the weight? For metrics based on the random forest, it is simplest to associate this weight with the RF proximity itself. In the specific case of the RF GAP proximity, these weights have the further compelling interpretation that they are exactly equal to the proportional contribution of each point (bond) in the training set to the RF prediction for the bond of interest.
However, this interpretation of the proximities is not available for any of the other metrics, obstructing a fair, like-for-like comparison among the various proximity-weighted KNN predictions for different metrics. 

\begin{figure}[H]
\centering

\begin{subfigure}{0.49\textwidth}

    \includegraphics[scale=0.45]{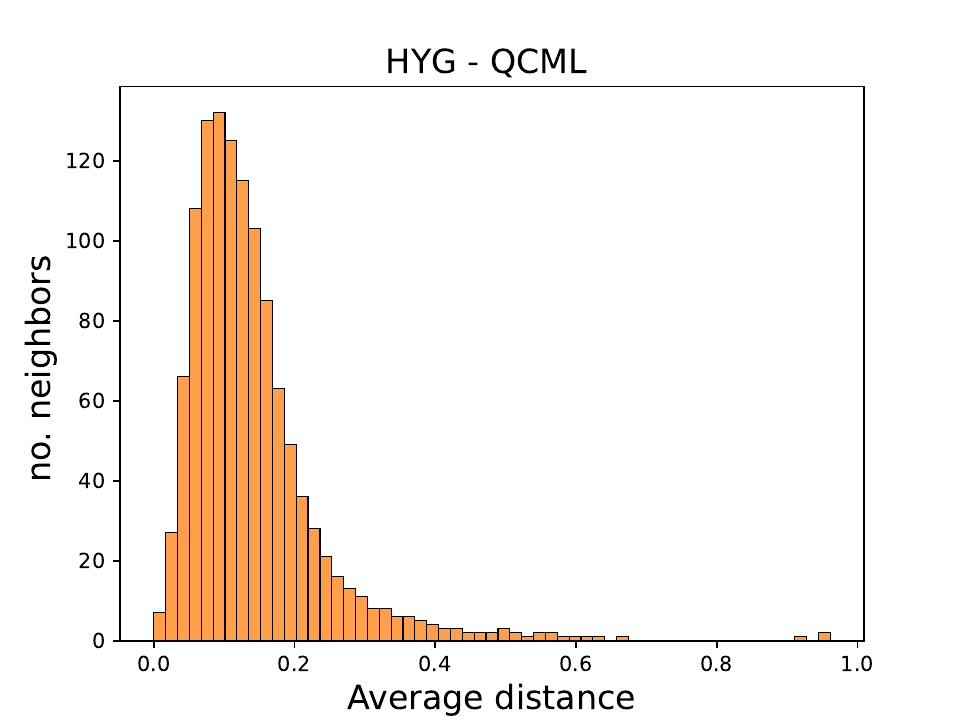}
    \caption{} 
\end{subfigure}
\begin{subfigure}{0.49\textwidth}

    \includegraphics[scale=0.45]{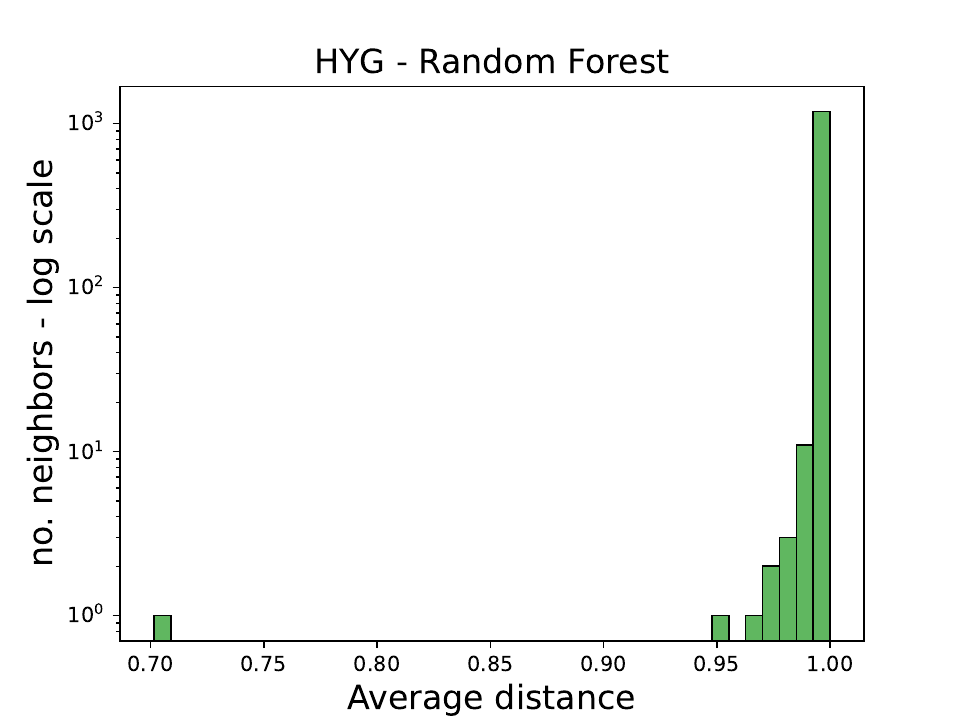}
    \caption{ }
\end{subfigure}

\medskip

\begin{subfigure}{0.49\textwidth}

    \includegraphics[scale=0.45]{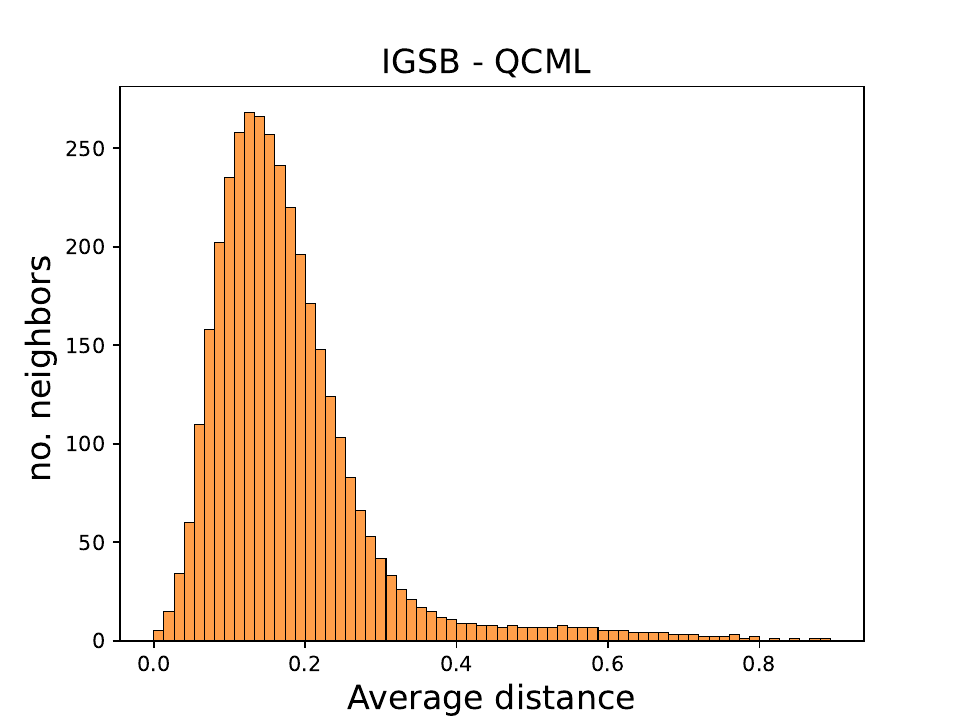}
        \caption{ }
\end{subfigure}
\begin{subfigure}{0.49\textwidth}

    \includegraphics[scale=0.45]{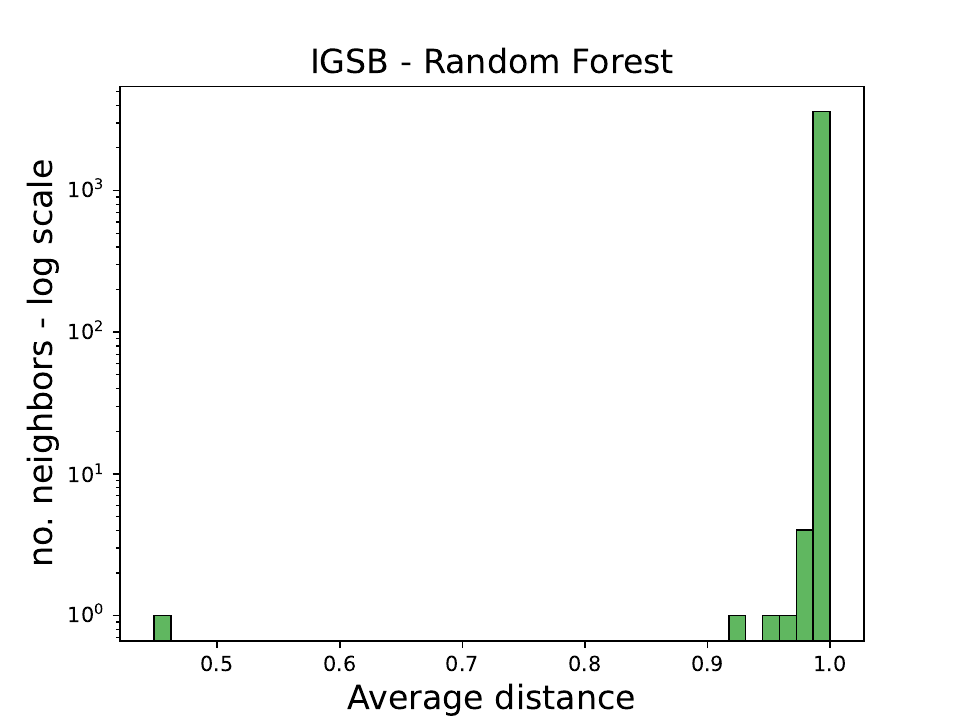}
    \caption{ }
\end{subfigure}

\caption{Average distance from a reference bond to its neighbors for both HYG and IGSB. The average is taken over all bonds in the data set. For Random Forest (GAP proximity), almost all neighbors are at a maximum distance from the reference bond. Note that for Random Forest the $y$-axis is in logarithmic scale. }
\label{fig:dist_hist}
\end{figure}

In principle, assuming similar yields, the question of whether another bond has sufficiently similar values for input features like rating, industry, etc. to use it as a substitute for the desired bond should be model-independent; on the other hand, basing the weights in the KNN average on model proximity makes these strongly model-dependent. This point is reinforced by the plots in Figure \ref{fig:dist_hist}, which show distributions of average distance for QCML and for the original RF proximity (where distance again is defined as 1 minus proximity). With the QCML metric, distances on the whole tend to be less than .5, while for RF, they tend for the most part to be greater than .9, entailing that for RF only a relatively small number of bonds contribute substantially to the proximity-weighted KNN prediction, while for QCML, a substantially broader region of neighbors contribute. This point is illustrated in a different way by the multi-dimensional scaling (MDS) plots of Section \ref{sec:MDS}, which demonstrate that under the QCML metric, points tend to be close together, while under the RF metric, they are typically far apart, and even more so for HYG than for IGSB bonds.

Given the element of arbitrariness that enters into the choice of a weighting scheme for computing the KNN prediction, when comparing performance of different metrics, we plot performance as a function of $k$ both without weighting, to show what the results look like in the absence of a weighting scheme, and also with weighting by proximity.

\section{Results}
\label{section:results}

Our main goal in this section is to evaluate the relative quality of the similarity matrices extracted from QCML, as compared with those extracted from the standard Euclidean metric, and from state-of-the-art supervised similarity learning with random forests.

\subsection{Target variable prediction}
We first test the QCML model on the regression task to verify its prediction performance. We repeated the test over 10 different 80/20 train-test splits of the data, where the data is randomly shuffled before each split. The regression metrics tracked are mean absolute percentage error (MAPE), mean absolute error (MAE), root mean squared error (RMSE) and $R^2$, The regression results for QCML are compared to that of linear regression as a baseline and with Random Forest. The results are shown in Table \ref{tab:regression_res}.

\begin{comment}
\begin{figure}[H]

\begin{subfigure}{\textwidth}

\centering
    \begin{tabular}{||c|ccc||}
    \hline
 \multicolumn{4}{|c|}{HYG - price prediction} \\
 \hline
          & $R^2$ & RMSE  &  MAPE \\
         \hline
         Linear Regression & 0.603 {\scriptsize  $\pm$ 0.073}  & 6.03 {\scriptsize  $\pm$ 0.7} & 4.31 {\scriptsize  $\pm$ 0.51}  \\
         Random Forest     & 0.606 {\scriptsize  $\pm$ 0.068}& 6.01 {\scriptsize  $\pm$ 0.9}  & 3.7 {\scriptsize  $\pm$ 0.66}  \\
         QCML              & 0.695 {\scriptsize  $\pm$ 0.073}  & 5.23 {\scriptsize  $\pm$ 0.92} & 3.02 {\scriptsize  $\pm$ 0.57} \\
\hline
\end{tabular}
\end{subfigure}

\medskip
\medskip

\begin{subfigure}{\textwidth}

\centering

    \begin{tabular}{||c|ccc||}
    \hline
 \multicolumn{4}{|c|}{IGSB - credit spread prediction } \\
 \hline
          & $R^2$ & RMSE  & MAPE \\
         \hline
         Linear Regression & 0.655 {\scriptsize  $\pm$ 0.056} & 27.5 {\scriptsize  $\pm$ 3.9} & 24.3 {\scriptsize  $\pm$ 2.4} \\
         Random Forest     & 0.717 {\scriptsize  $\pm$ 0.06} & 24.8 {\scriptsize  $\pm$ 3.6}  & 20.35 {\scriptsize  $\pm$ 2.3} \\
         QCML              & 0.727 {\scriptsize  $\pm$ 0.055} & 24.5 {\scriptsize  $\pm$ 4.1} & 20.35 {\scriptsize  $\pm$ 2.2} \\
\hline
\end{tabular}

\end{subfigure}

\caption{Average test set metrics and standard deviations for target variable predictions on both HYG and IGSB data sets. The average and standard deviations are taken over 50 different 80/20 train/test splits of the data.    }
\label{tab:regression_res}

\end{figure}
\end{comment}

\begin{table}[H]

\begin{subtable}{\textwidth}

\centering
    \begin{tabular}{||c|cccc||}
    \hline
 \multicolumn{4}{|c|}{HYG - yield prediction} \\
 \hline
          & MAPE & MAE & RMSE  & $R^2$ \\
         \hline
         Linear Regression & .129 {\scriptsize  $\pm$ .007}  & 1.06 {\scriptsize  $\pm$ .08} & 1.86 {\scriptsize  $\pm$ .19} & .59 {\scriptsize  $\pm$ .06} \\
         Random Forest  & .11 {\scriptsize  $\pm$ .007}& .93 {\scriptsize  $\pm$ .08}  & 1.80 {\scriptsize  $\pm$ .19} & .62 {\scriptsize  $\pm$ .04}\\
         QCML & \textbf{.09} {\scriptsize  $\pm$ .01}  & \textbf{.79} {\scriptsize $\pm$ .06} & \textbf{1.73} {\scriptsize  $\pm$ .17} & \textbf{.65} {\scriptsize  $\pm$ .07} \\
\hline
\end{tabular}
\end{subtable}

\medskip
\medskip

\begin{subtable}{\textwidth}
\centering
    \begin{tabular}{||c|cccc||}
    \hline
 \multicolumn{4}{|c|}{IGSB - credit spread prediction} \\
 \hline
          & MAPE & MAE & RMSE  & $R^2$ \\
         \hline
        Linear Regression & .26 {\scriptsize  $\pm$ .03}  & 15.90 {\scriptsize  $\pm$ .48} & 22.83 {\scriptsize  $\pm$ 1.40} & .69 {\scriptsize  $\pm$ .01} \\
         Random Forest & .25 {\scriptsize  $\pm$ .04}& \textbf{13.74} {\scriptsize  $\pm$ .47}  &  \textbf{20.30} {\scriptsize  $\pm$ 1.28} & \textbf{.76} {\scriptsize  $\pm$ .02}\\
         QCML              & \textbf{.23} {\scriptsize  $\pm$ .05}  & 13.87 {\scriptsize  $\pm$ .49} & 20.78 {\scriptsize $\pm$ 1.08} & .74 {\scriptsize  $\pm$ .02} \\
\hline
\end{tabular}
\end{subtable}

\caption{Average test set metrics and standard deviations for target variable predictions on both HYG and IGSB data sets. The average and standard deviations are taken over 10 different 80/20 train/test splits of the data.}
\label{tab:regression_res}
\end{table}

\begin{table}[H]

\begin{comment}
\begin{subfigure}{\textwidth}

\centering
    \begin{tabular}{||c|cccc||}
    \hline
 \multicolumn{4}{|c|}{California Housing} \\
 \hline
          & MAPE & MAE & RMSE  & $R^2$ \\
         \hline
         Linear Regression & . {\scriptsize  $\pm$ }  &  {\scriptsize  $\pm$ } &  {\scriptsize  $\pm$ } &  {\scriptsize  $\pm$ } \\
         Random Forest  &  {\scriptsize  $\pm$ }&  {\scriptsize  $\pm$ }  &  {\scriptsize  $\pm$ } & {\scriptsize  $\pm$ }\\
         QCML & {\scriptsize  $\pm$ }  & {\scriptsize $\pm$ } & {\scriptsize  $\pm$ } &  {\scriptsize  $\pm$ } \\
\hline
\end{tabular}
\end{subfigure}
\end{comment}

\begin{subtable}{\textwidth}
    \centering
    \begin{tabular}{||c|cccc||}
    \hline
    \multicolumn{4}{|c|}{California Housing} \\
    \hline
          & MAPE & MAE & RMSE  & $R^2$ \\
         \hline
         Linear Regression & .28 {\scriptsize  $\pm$ .02}  & .50 {\scriptsize  $\pm$ .02} & .74 {\scriptsize  $\pm$ .05} &  .58 {\scriptsize  $\pm$ .07} \\
         Random Forest  & .24 {\scriptsize  $\pm$ .02}& \textbf{.38} {\scriptsize  $\pm$ .01}  & \textbf{.55} {\scriptsize  $\pm$ .02} & \textbf{.77} {\scriptsize  $\pm$ .01}\\
         QCML & \textbf{.21} {\scriptsize  $\pm$ .01}  & \textbf{.38} {\scriptsize $\pm$ .02} & .56 {\scriptsize $\pm$ .04} & .76 {\scriptsize  $\pm$ .03} \\
\hline
\end{tabular}
\end{subtable}

\medskip
\medskip

\begin{subtable}{\textwidth}

\centering
    \begin{tabular}{||c|cccc||}
    \hline
 \multicolumn{4}{|c|}{Student Performance} \\
 \hline
          & MAPE & MAE & RMSE  & $R^2$ \\
         \hline
         Linear Regression & $\infty$   & .57 {\scriptsize  $\pm$ .03} & .80 {\scriptsize  $\pm$ .05} & .58 {\scriptsize  $\pm$ .05} \\
         Random Forest & $\infty$  & .53 {\scriptsize $\pm$ .02}  & \textbf{.75} {\scriptsize  $\pm$ .04} & \textbf{.62} {\scriptsize  $\pm$ .03}\\
         QCML & $\infty$ & \textbf{.41} {\scriptsize $\pm$ .02} & \textbf{.75} {\scriptsize $\pm$ .05} & \textbf{.62} {\scriptsize  $\pm$ .05} \\
\hline
\end{tabular}
\end{subtable}

\medskip
\medskip

\begin{subtable}{\textwidth}

\centering
    \begin{tabular}{||c|cccc||}
    \hline
 \multicolumn{4}{|c|}{Diabetes} \\
 \hline
          & MAPE & MAE & RMSE  & $R^2$ \\
         \hline
         Linear Regression & \textbf{.38} {\scriptsize  $\pm$ .04}  & \textbf{43.43 }{\scriptsize  $\pm$ 3.07} & \textbf{54.47} {\scriptsize  $\pm$ 3.61} & \textbf{.45} {\scriptsize  $\pm$ .08} \\
         Random Forest & .39  {\scriptsize  $\pm$ .03}& 45.34 {\scriptsize  $\pm$ 2.13}  & 56.24 {\scriptsize  $\pm$ 2.88} & .42{\scriptsize  $\pm$ .07}\\
         QCML & .39{\scriptsize  $\pm$ .03}  & 48.05{\scriptsize $\pm$ 3.67} & 60.52 {\scriptsize  $\pm$ 5.53} & .32 {\scriptsize  $\pm$ .16} \\
\hline
\end{tabular}
\end{subtable}

\medskip
\medskip

\begin{comment}
\begin{subfigure}{\textwidth}

\centering
    \begin{tabular}{||c|cccc||}
    \hline
 \multicolumn{4}{|c|}{Air Quality} \\
 \hline
          & MAPE & MAE & RMSE  & $R^2$ \\
         \hline
         Linear Regression & 15.58 {\scriptsize $\pm$ 2.60}  & 30.03 {\scriptsize  $\pm$ 2.01} &  58.65 {\scriptsize  $\pm$ 3.22} & .44 {\scriptsize  $\pm$ .054} \\
         Random Forest & 12.13 {\scriptsize  $\pm$ 2.92}&  21.20 {\scriptsize $\pm$ 2.41}  &  57.19 {\scriptsize  $\pm$ 3.85} & .46 {\scriptsize  $\pm$ .07}\\
         QCML & 11.46 {\scriptsize  $\pm$ 2.99}  & 18.34 {\scriptsize $\pm$ 2.99} & 60.13{\scriptsize $\pm$ 3.76} & .41 {\scriptsize  $\pm$ .07} \\
\hline
\end{tabular}
\end{subfigure}
\end{comment}

\medskip
\medskip
\begin{comment}
\begin{subfigure}{\textwidth}

\centering
    \begin{tabular}{||c|cccc||}
    \hline
 \multicolumn{4}{|c|}{Bike Share} \\
 \hline
          & MAPE & MAE & RMSE  & $R^2$ \\
         \hline
         Linear Regression & 7.04e-5 {\scriptsize  $\pm$ 1.14e-5}  &  .21 {\scriptsize $\pm$ .04} &  .26{\scriptsize  $\pm$ .05} &  {.999999982 \scriptsize  $\pm$ 7e-9} \\
         Random Forest  & .08 {\scriptsize  $\pm$ .01}& 216.25 {\scriptsize $\pm$ 22.61}  & 319.96 {\scriptsize $\pm$ 33.77} & .97 {\scriptsize  $\pm$ .01}\\
         QCML & .02{\scriptsize  $\pm$ .00}  & 47.73{\scriptsize $\pm$ 9.74} & 85.00{\scriptsize $\pm$ 19.57} &  {.998 \scriptsize  $\pm$ .001} \\
\hline
\end{tabular}
\end{subfigure}
\end{comment}
\caption{Out-of-sample performance metrics for linear regression, random forest regression, and QCML regression on three well-known public datasets. Note that the MAPE metric for the Student Performance dataset diverges since the target, representing grades, consists of integers including $0$; MAPE is not a meaningful metric in this case since it is only differences between grades, rather than there absolute scale, that has meaning here.}

\end{table}

Our main concern in this article is to evaluate the similarity matrices extracted from regression models like QCML or RF, rather than to evaluate the relative performance of the models themselves as predictors of a given target variable. Nevertheless, as we will see in the examples below, the model's performance according to traditional evaluation metrics for regression directly impacts the quality of the metric extracted from that model, as measured by the KNN method described above. 

For example, in the case of HYG, we see that the superior performance of QCML as a predictor of yield, shown in the upper box of Table \ref{tab:regression_res}, translates also into better results for the associated similarity metric, shown in Figure \ref{fig:KNN_results} (a), (b). By contrast, in the case of IGSB, the fact that the performance of QCML as a predictor of spread is relatively closer to that of RF translates into relatively more comparable performance for associated distance metrics, particularly in the case of proximity weighting shown in Figure \ref{fig:KNN_results} (d); for the unweighted KNN curve in Figure \ref{fig:KNN_results} (c), QCML still does better for the most part. 

Among the public datasets, we see that for the California Housing dataset, where QCML and RF perform on par, the associated QCML metric performs similarly but slightly better than the other metrics for the unweighted case; see Figure \ref{fig:KNN_results_public} (a). On the Student Performance dataset, where the QCML regressor performs better than or on par with RF and Linear Regression, the QCML metric outperforms both Euclidean and RF-GAP metrics across all values of $k$ plotted. In the Diabetes dataset, where QCML under-performs as a regression model, it performs roughly on par, and for small $k$ slightly better than, RF-GAP and Euclidean metrics; see Figure \ref{fig:KNN_results_public} (e), (f).

It is interesting to note that linear regression works
well for the Diabetes dataset, outperforming both RF and QCML. Linear regression works well with a small dataset (442 data instances in the case of Diabetes) because it is less prone to overfitting than random forests, which can easily
overfit if not correctly tuned.  Random forests shine when there are
complex interactions and non-linear relationships between variables.
If such interactions do not exist, then the added complexity of a
random forest may not provide any benefit and could even harm
performance.  Breiman \cite{breiman-cutler-blog} introduces random forests and discusses
their advantages over linear models, emphasizing that they are
particularly powerful when data contains strong interactions and
non-linearities. Conversely, simpler models like linear regression can
be just as effective if no such complexities exist. A similar discussion can be applied to QCML. As shown in \cite{CandeloriEtAl}, the quantum model of the data manifold produced by QCML tends to be a compact, highly non-linear manifold, which is not optimal when approximating (unbounded) linear manifolds.

\subsection{Similarity measure evaluation}

Next, we compute the similarity matrices and apply the KNN method to evaluate each metric. Three distance metrics are evaluated: 1) standard (unsupervised) Euclidean, 2) random forest GAP 3) QCML. For each distance metric we report the test MAPE or MAE (in cases where the MAPE diverges) of a KNN regressor for a range of neighbors $K=1, \ldots, 100$, first in the case where the KNN average is unweighted, and then in the case where the KNN average is weighted by proximity. The KNN method for evaluating similarity is performed for bonds in the HYG and IGSB indices, as well as for three well-known public datasets. 

\begin{figure}[H]
\centering

\begin{subfigure}{0.49\textwidth}

    \includegraphics[scale=0.3]{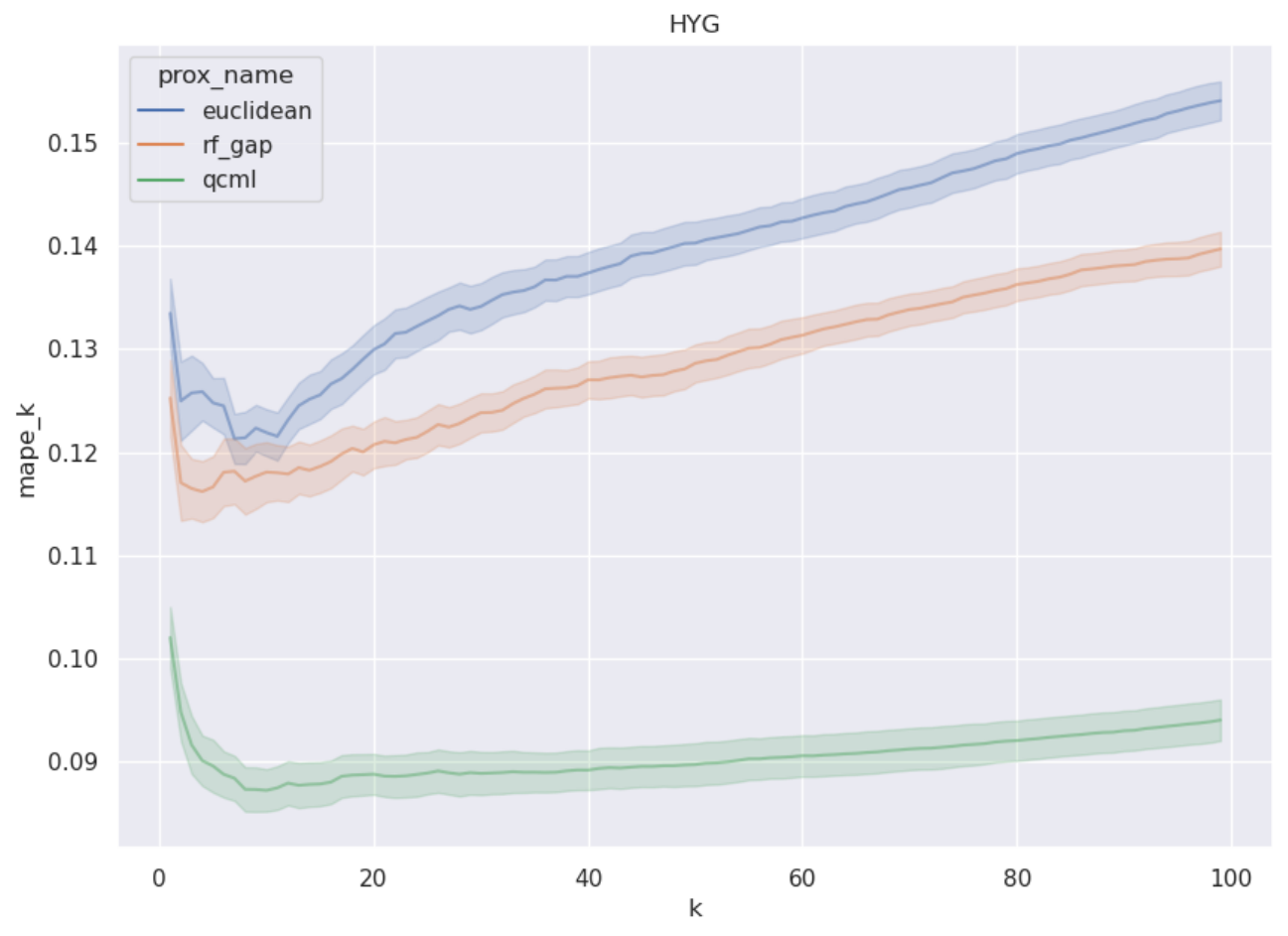}
    \caption{} 
\end{subfigure}
\begin{subfigure}{0.49\textwidth}

    \includegraphics[scale=0.3]{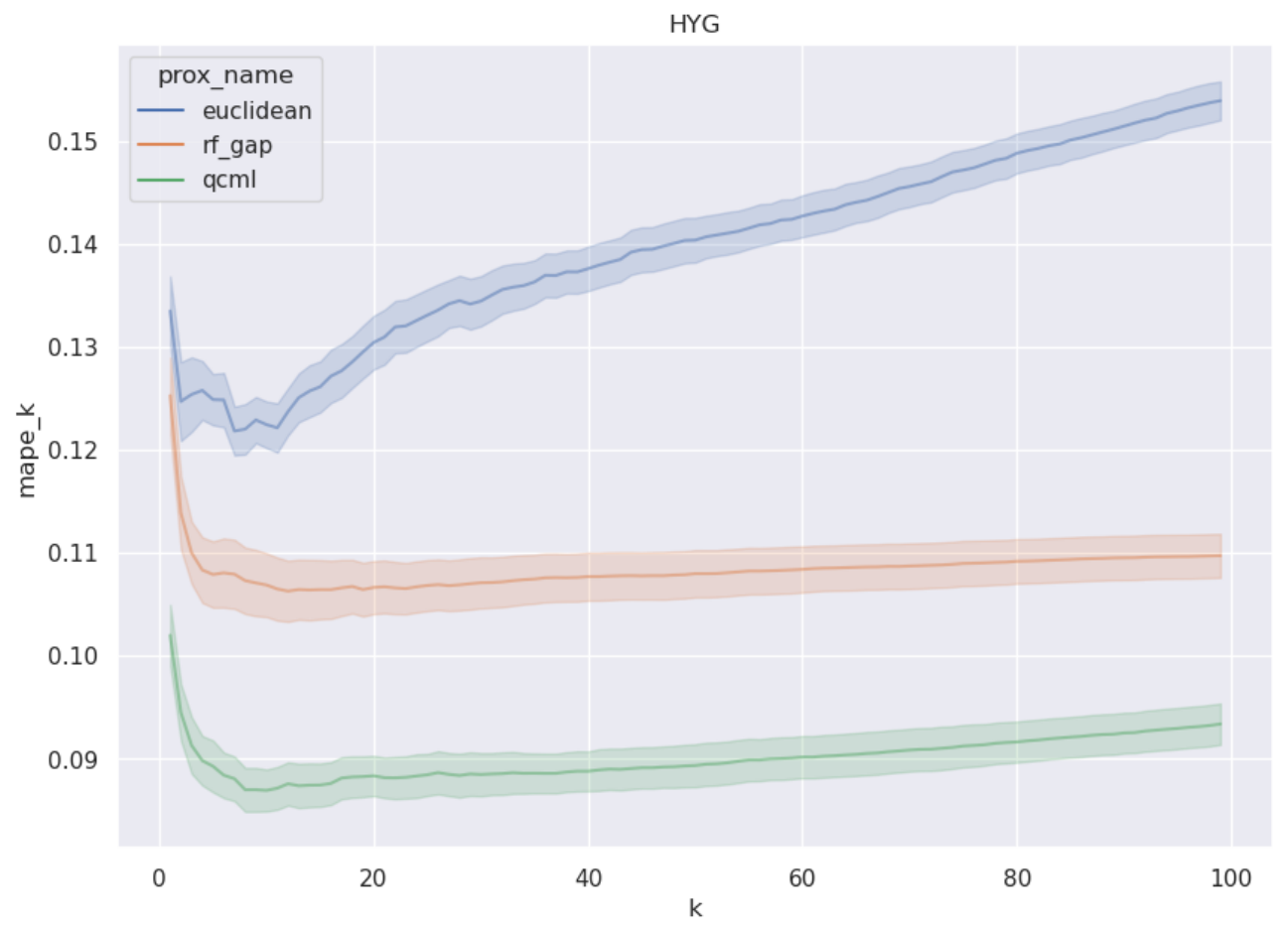}
    \caption{ }
\end{subfigure}
\begin{subfigure}{0.49\textwidth}

    \includegraphics[scale=0.3]{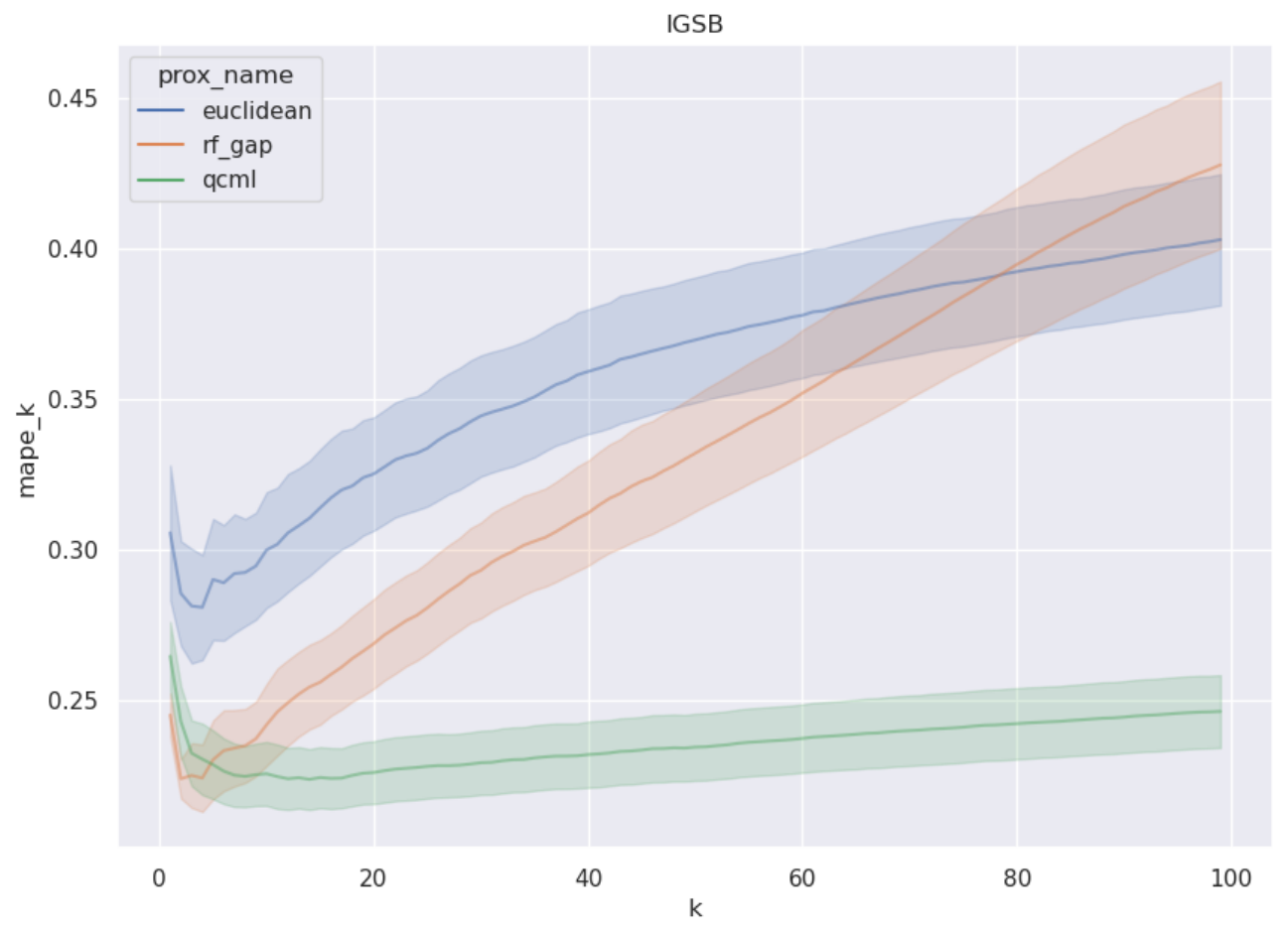}
    \caption{ }
\end{subfigure}
\begin{subfigure}{0.49\textwidth}

    \includegraphics[scale=0.3]{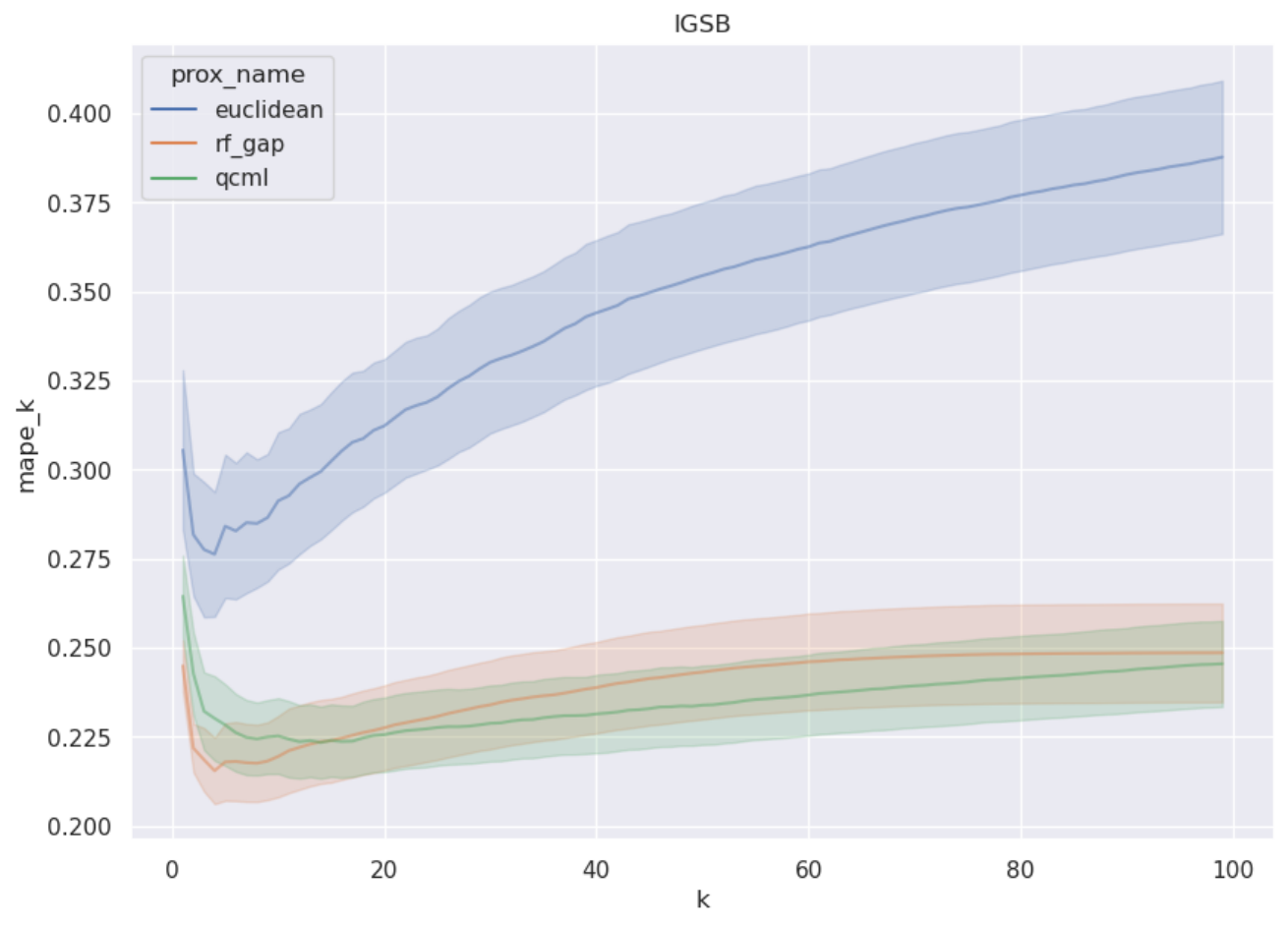}
    \caption{ }
\end{subfigure}

\caption{Average MAPE for K-Nearest-Neighbors (KNN) regression over 10 train/test splits of the data, for both  HYG (upper plots) and  IGSB (lower plots). The bands represent standard error in estimation of the mean for each $k$. In the plots on the left, a) and c), the KNN prediction is computed as a simple average of the training targets for the k nearest neighbors. In the plots on the right, b) and d), the KNN prediction is computed as a proximity-weighted average of the $k$-nearest neighbors.}
\label{fig:KNN_results}
\end{figure}

\begin{figure}[H]
\centering

\begin{subfigure}{0.49\textwidth}

    \includegraphics[scale=0.3]{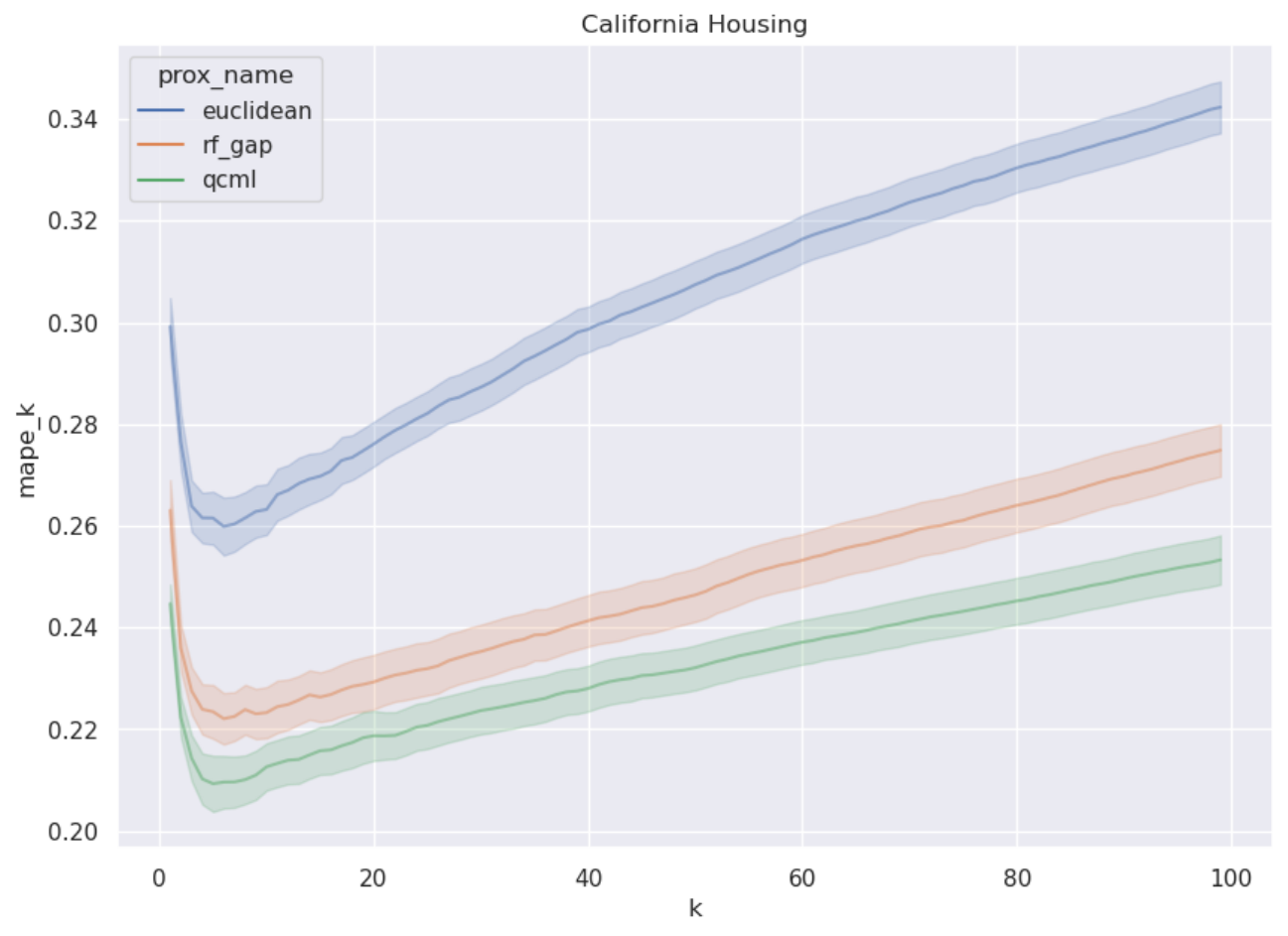}
    \caption{} 
\end{subfigure}
\begin{subfigure}{0.49\textwidth}

    \includegraphics[scale=0.3]{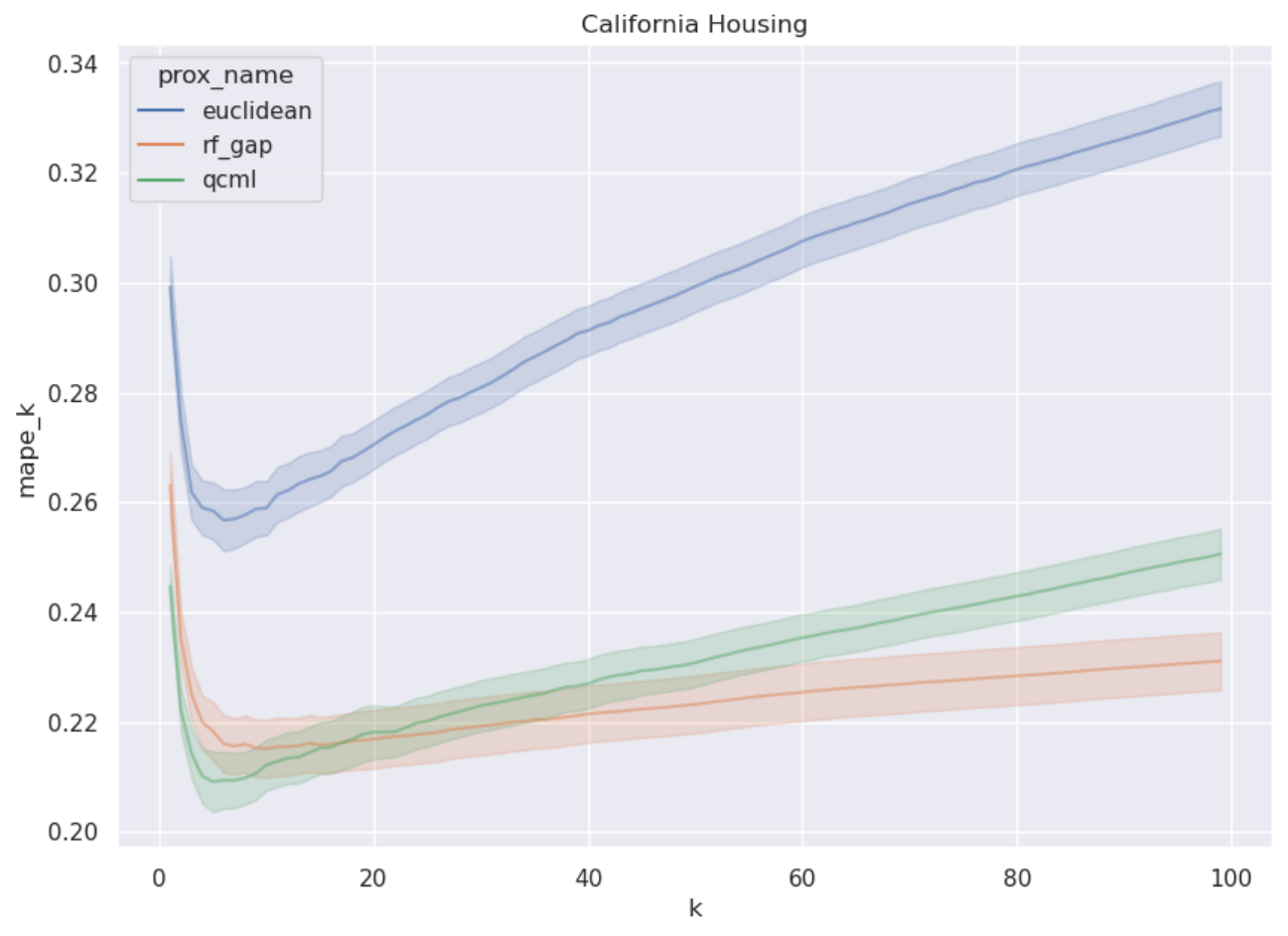}
    \caption{ }
\end{subfigure}

\begin{subfigure}{0.49\textwidth}
    \includegraphics[scale=0.3]{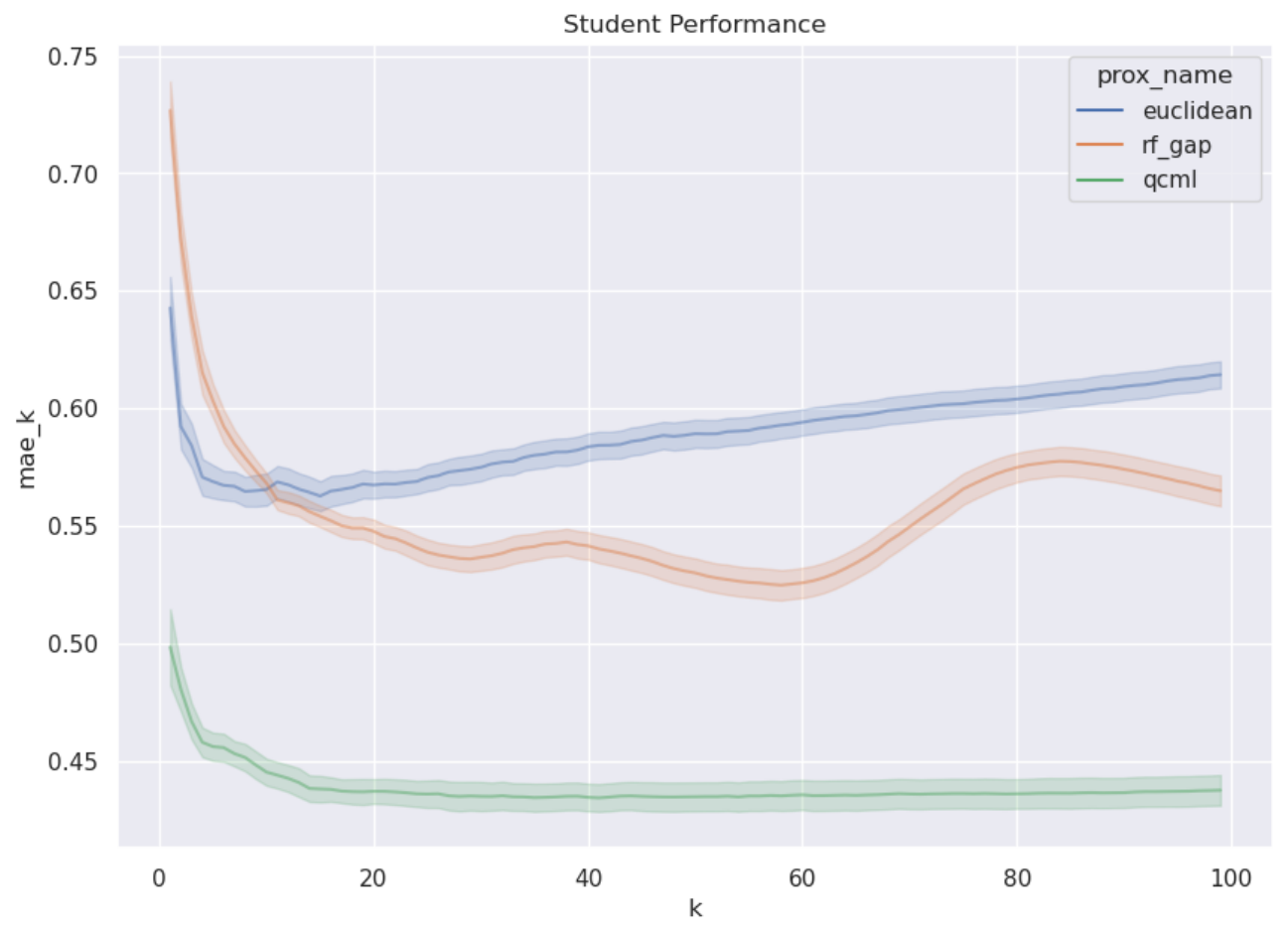}
    \caption{ }
\end{subfigure}
\begin{subfigure}{0.49\textwidth}

    \includegraphics[scale=0.3]{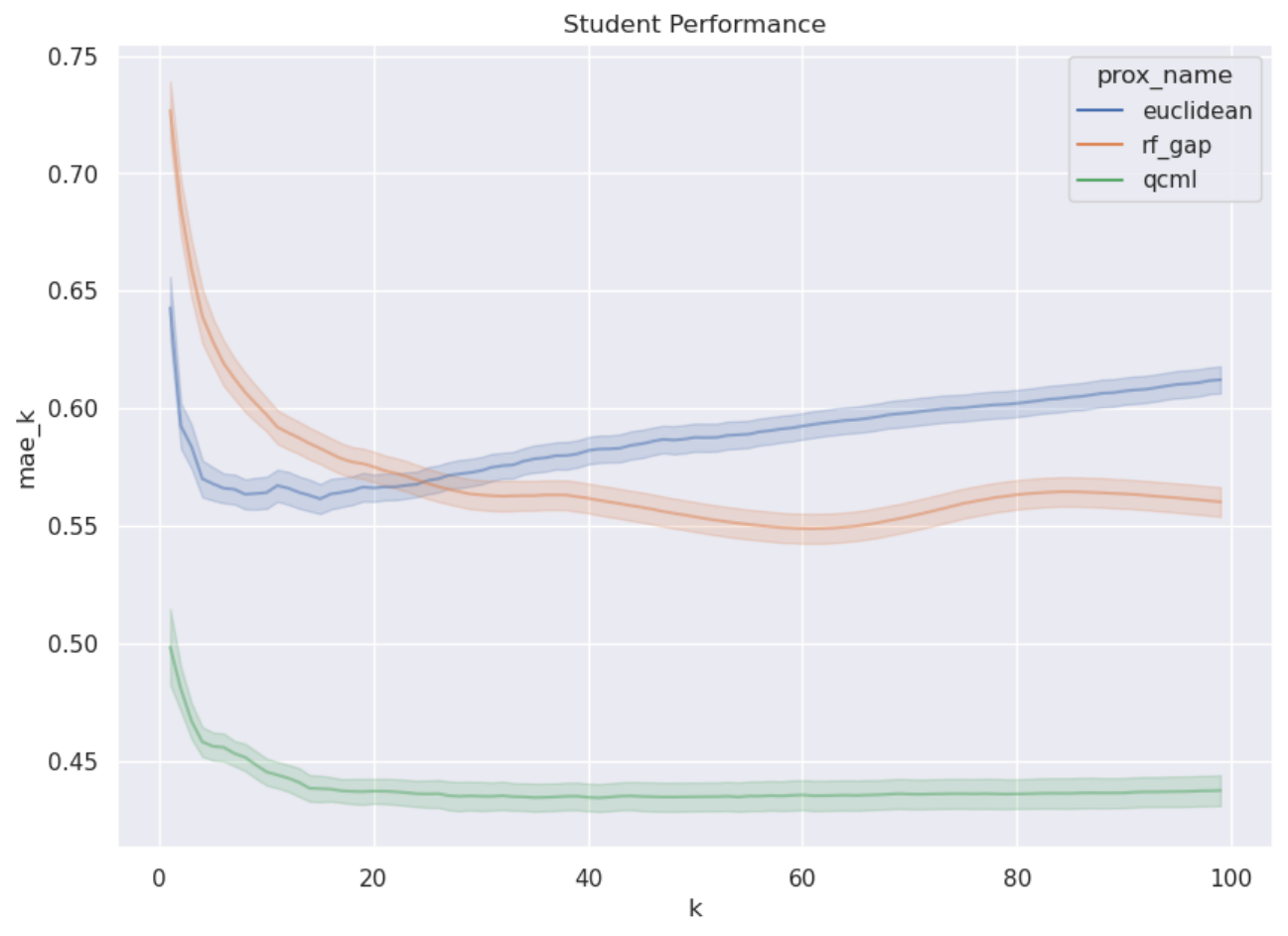}
    \caption{ }
\end{subfigure}

\begin{subfigure}{0.49\textwidth}
    \includegraphics[scale=0.3]{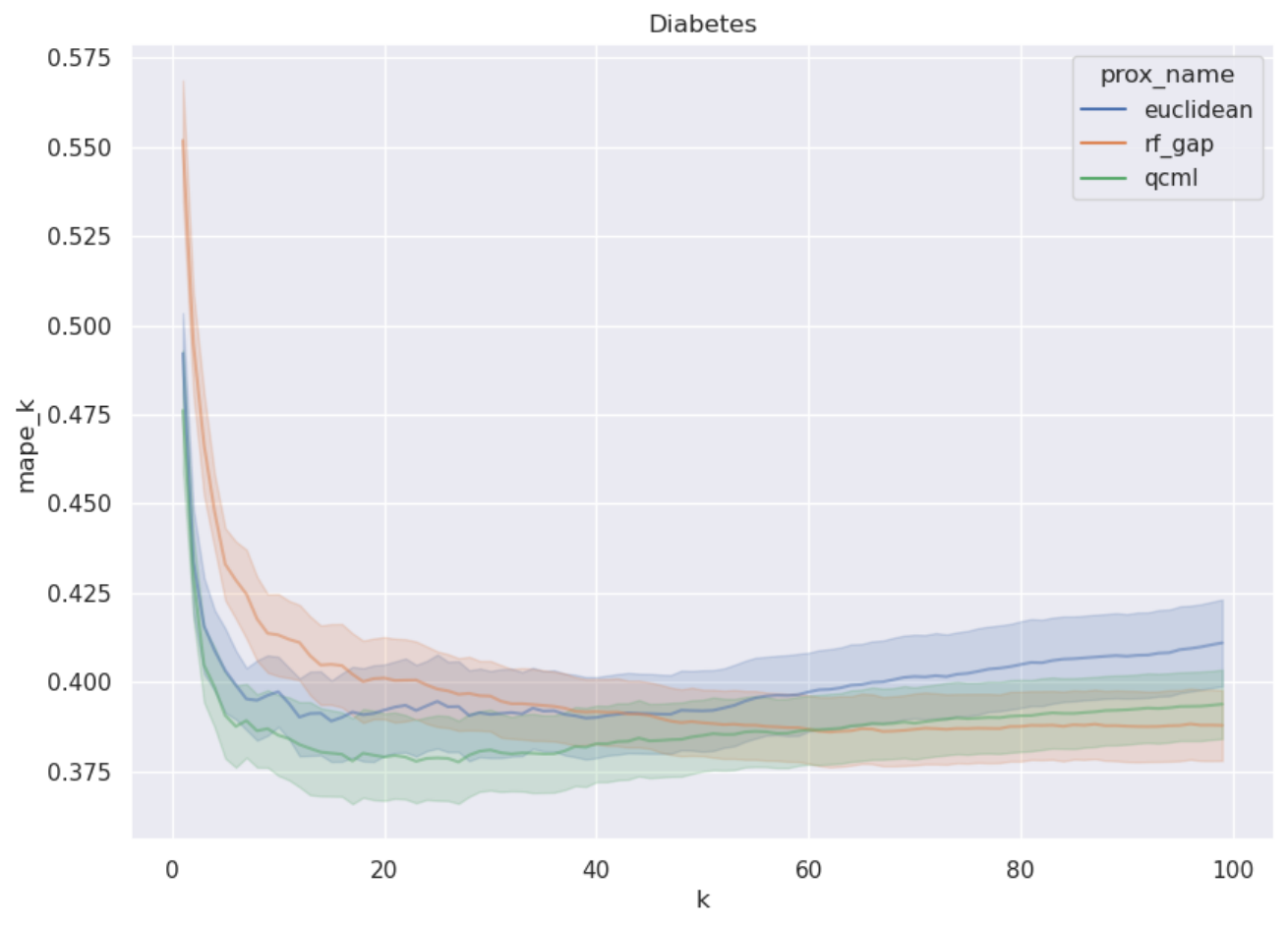}
    \caption{ }
\end{subfigure}
\begin{subfigure}{0.49\textwidth}

    \includegraphics[scale=0.3]{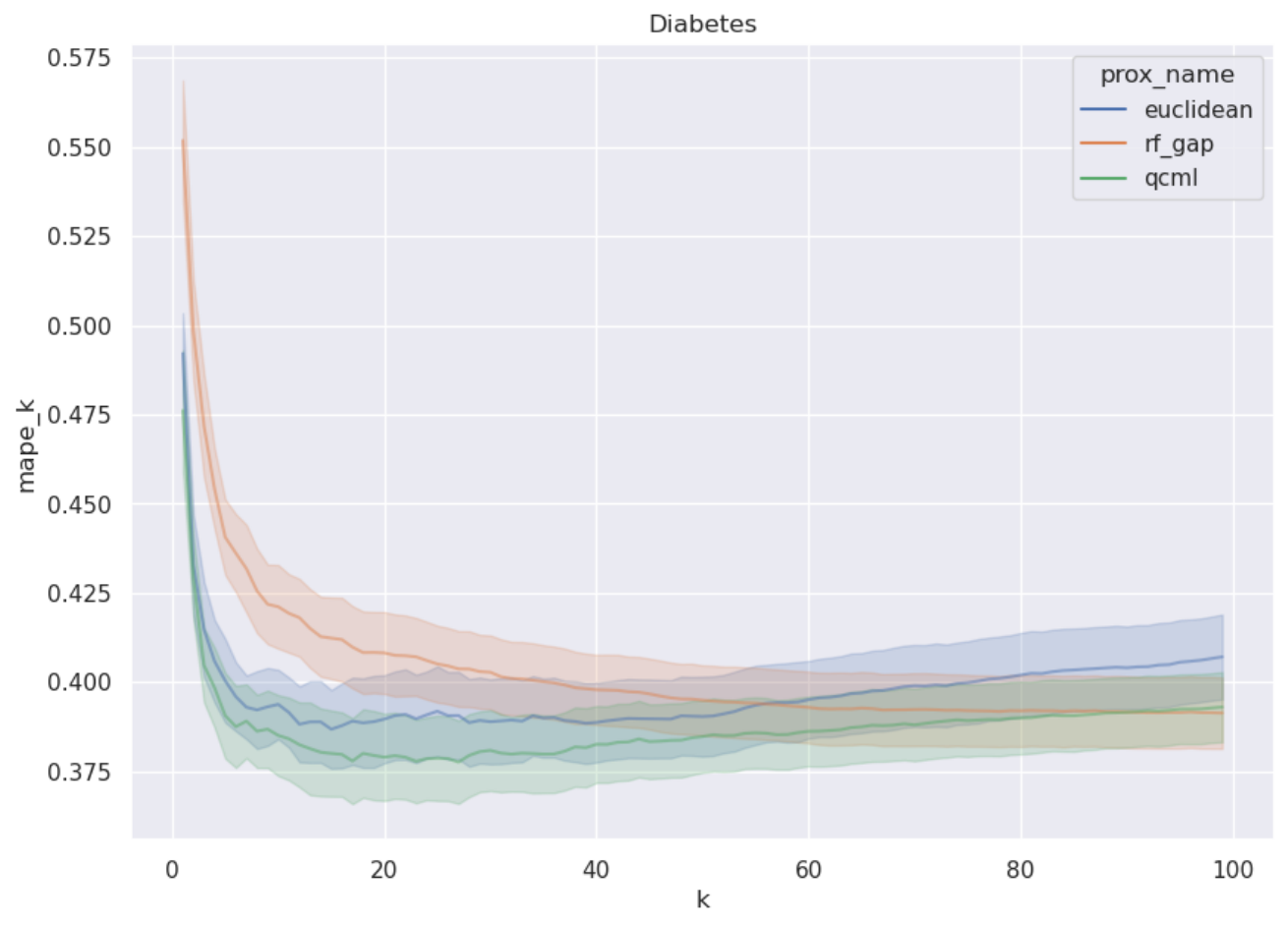}
    \caption{ }
\end{subfigure}

\begin{comment}
\begin{subfigure}{0.49\textwidth}
    \includegraphics[scale=0.35]{diabetes/mape_unweighted.png}
    \caption{ }
\end{subfigure}
\begin{subfigure}{0.49\textwidth}

    \includegraphics[scale=0.35]{diabetes/mape_proxweighted.png}
    \caption{ }
\end{subfigure}
\begin{subfigure}{0.49\textwidth}

    \includegraphics[scale=0.35]{airquality/mape_unweighted.png}
    \caption{ }
\end{subfigure}
\begin{subfigure}{0.49\textwidth}

    \includegraphics[scale=0.35]{airquality/mape_proxweighted.png}
    \caption{ }
    \end{subfigure}

\begin{subfigure}{0.49\textwidth}
   \includegraphics[scale=0.35]{bikesharing/mape_unweighted.png}
    \caption{ }
\end{subfigure}
\begin{subfigure}{0.49\textwidth}
    \includegraphics[scale=0.35]{bikesharing/mape_proxweighted.png}
    \caption{ }
\end{subfigure}
\end{comment}

\caption{Average out-of-sample MAPE or MAE as a function of k for k-nearest-neighbors regression over 10 train/test splits of the data, for three well-known public datasets. Each row corresponds to a distinct dataset; plots on the left correspond to unweighted KNN and plots on the right correspond to proximity-weighted KNN. The bands represent standard error in estimation of the mean MAPE or MAE over the 10 splits. MAPE is used by default; however, the target variable in the Student Performance dataset consists of integer grade values from 0 to 4; because the presence of 0 as one of these values causes the computation of MAPE to diverge, we use MAE for this dataset instead.}
\label{fig:KNN_results_public}

\end{figure}

The results of the corporate bond analysis are shown in Figure \ref{fig:KNN_results}. For the public datasets, they are shown in Figure \ref{fig:KNN_results_public}. 

For the corporate bonds dataset we see in Figure \ref{fig:KNN_results} that for HYG, the QCML metric outperforms other metrics for both the unweighted and proximity-weighted methods of KNN regression. For IGSB, we see that the QCML metric outperforms again for the unweighted KNN regression, and performs comparably, if slightly worse, for the proximity-weighted KNN regression. As we discuss below, RF-based proximities tend to be at or near the minimum value of 0 for all but a relatively small proportion of training points; as a result, these proximities tend to place more weight on a smaller number of nearest neighbors than do QCML proximities, which have support over a larger proportion of the training set. In this respect, proximity-weighting inherently favors the RF-based proximities, as it counts only a small number of nearest neighbors that are more likely to have yields close to that of the test point.

For public datasets, we see in Figure \ref{fig:KNN_results_public} that the QCML metric again outperforms both RF-GAP and Euclidean metrics using the unweighted method of KNN regression. For the proximity-weighted method, it outperforms these methods up to $k=15$, but for larger values of $k$, the RF-GAP metric does better. On the Student Performance dataset, the QCML metric performs uniformly better than both RF-GAP and Euclidean metrics across the range of $k$ plotted. On the Diabetes dataset, QCML performs comparably to both RF-GAP and Euclidean metrics, performing slightly better than both for $k$ less than $50$ or so, and slightly worse for larger values of $k$.

\begin{comment}
\begin{figure}[H]
\centering

\begin{subfigure}{0.49\textwidth}

    \includegraphics[scale=0.35]{hyg_knn_plot.pdf}
    \caption{} 
\end{subfigure}
\begin{subfigure}{0.49\textwidth}

    \includegraphics[scale=0.35]{ig_knn_plot.pdf}
    \caption{ }
\end{subfigure}

\medskip

\begin{subfigure}{0.49\textwidth}

    \includegraphics[scale=0.35]{hyg_knn_plot_unweighted.pdf}
    \caption{} 
\end{subfigure}
\begin{subfigure}{0.49\textwidth}

    \includegraphics[scale=0.35]{ig_knn_plot_unweighted.pdf}
    \caption{ }
\end{subfigure}

\caption{Average MAPE for K-nearest-neighbors regression over 50 train/test splits of the data, for both  HYG and  IGSB. The bands represent 95\% confidence intervals. }
\label{fig:KNN_results}
\end{figure}
\end{comment}

% Note first that both QCML and Random Forest have much better performance than the standard Euclidean distance. This is expected, since both QCML and Random Forest distances are supervised, while the Euclidean metric is unsupervised. For the IGSB data set, the optimal performance of QCML and Random Forest is essentially equivalent, and it occurs at around $k=5$ neighbors (Figure \ref{fig:KNN_results}-b). For high-yield bonds (Figure \ref{fig:KNN_results}-a), QCML has a statistically significant advantage pretty much over the entire range of neighbors. For both data sets, as the number of neighbors increases the performance of Random Forest deteriorates faster than QCML, indicating a better ability of QCML to capture global, long-range similarities between data.  Indeed, the optimal number of neighbors for Random Forest proximity should be approximately equal to the average number of samples contained in each leaf. Unconstrained, this average number tends to be low, especially for sparse data such as high-yield bonds that exhibit large dispersion of the target variable. While it is possible during training to impose a minimum number of samples per leaf, ensuring that longer-range similarities are captured by Random Forest proximity, this constraint degrades the overall performance (see Figure \ref{fig:knn_leafs}). In contrast, the QCML model achieves maximum performance while at the same time preserving long-range similarities in the data. 

The most significant difference in performance between RF and QCML can be found in HY bonds (Figure \ref{fig:KNN_results} a-b). In binary classification, random forests can perform poorly on
imbalanced datasets, particularly when the minority class is
underrepresented. RFs are biased toward the majority class since they
minimize overall error, meaning they tend to predict the majority
class more often, which leads to a lower error rate. This results in
poor recall (sensitivity) for the minority class. In addition, since
random forests are built using decision trees, and decision trees tend
to favor classes with higher frequencies, the majority class gets more
influence in the splits. Since RF uses bootstrap sampling to train
individual trees, low minority classes are not represented in
bootstrapping; if the dataset is imbalanced, many bootstrapped samples
will contain very few instances of the minority class, leading to weak
predictive performance for those cases. In RF regression (RFR), imbalanced data can still cause issues, but
the challenges differ a bit from classification. Since RFR takes the
average of predictions from multiple decision trees, if one part of
the dataset dominates, the model’s predictions may be biased toward
the majority region of the target variable.  The model may struggle to
predict extreme values if the target variable is highly skewed because
few trees are trained on those cases. If most trees are trained on
similar value ranges (due to imbalanced data), the overall variance of
predictions is reduced, making the model less flexible in capturing
minority patterns. Consequently, RFR struggles with imbalanced target
distributions because it averages predictions, leading to poor
performance on rare values. In contrast, as we illustrate in the next Section, QCML is well-suited to handle imbalanced target distributions, since its predictions are based on a faithful representation of the underlying data manifold, including possibly sparse regions of the data.

\section{Visualizing bond similarities}
\label{sec:MDS}

\begin{figure}[H]
\centering
\begin{subfigure}{0.49\textwidth}
    \includegraphics[scale=0.6]{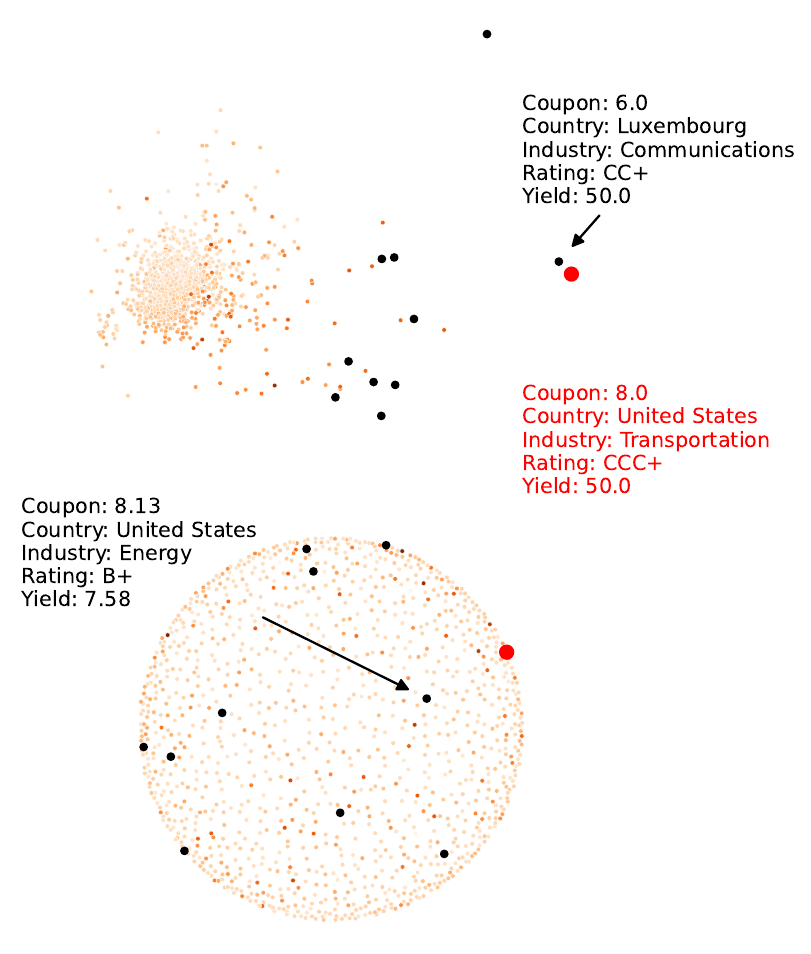}
    \caption{} 
\end{subfigure}
\begin{subfigure}{0.49\textwidth}
    \includegraphics[scale=0.6]{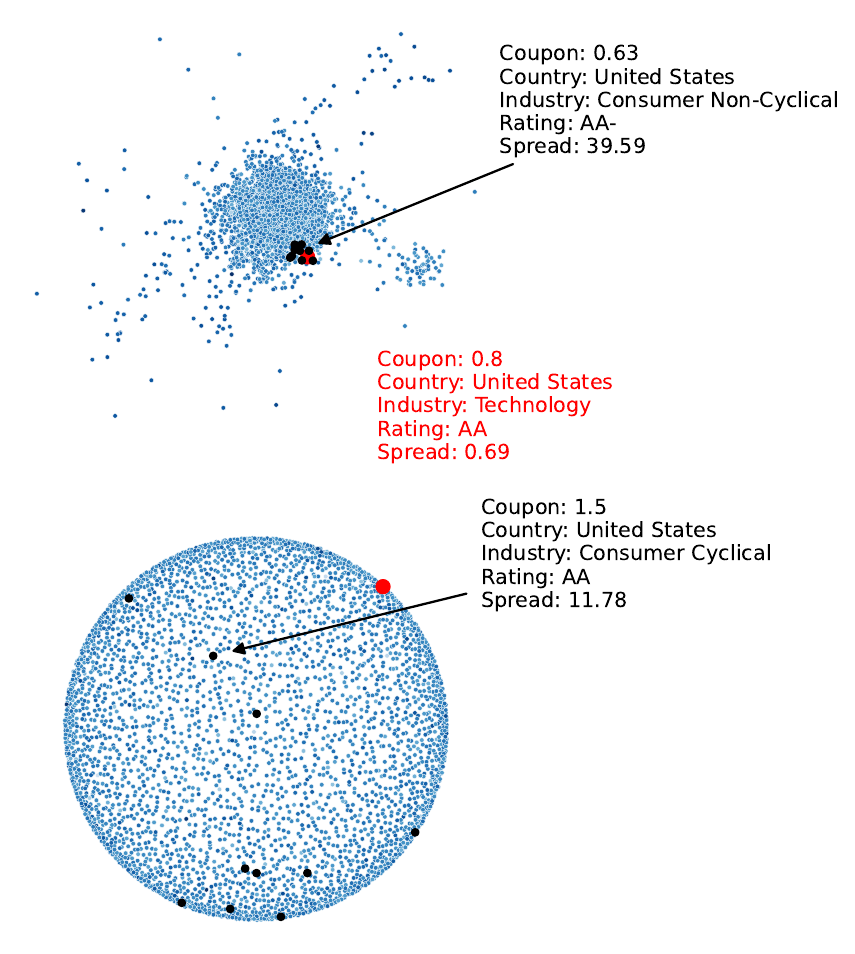}
    \caption{ }
\end{subfigure}

\caption{Multi-dimensional scaling visualization of both QCML (top) and Random Forest GAP proximities (bottom), for both (a) HYG and (b) IGSB. Darker colors represent higher yields. In all plots, a reference bond is displayed in red and the 10 closest neighbors are displayed in black. }
\label{fig:bond_plots}
\end{figure}

In this final section, we visualize both QCML and Random Forest (GAP) proximity metrics using multi-dimensional scaling (MDS). This is a dimension-reduction technique that can be used to visualize high-dimensional data in two dimensions. In a nutshell, MDS finds a mapping of the high-dimensional data into two dimensions to minimize the matrix norm of the difference between the distance matrix of the original data and that of the two-dimensional transformation. The technique can be applied using an arbitrary distance matrix, not necessarily corresponding to Euclidean distance, and is therefore suitable for visualizing both QCML and Random Forest GAP proximities. Using MDS we plotted the high-dimensional bond data for both HYG and IGSB in two dimensions, as shown in Figure \ref{fig:bond_plots}. 

While the plots in two-dimensional MDS target space do not offer a completely faithful representation of the distances between bonds, they help illustrate qualitatively some of the differences between QCML and Random Forest (GAP) proximities. For example, MDS is able to capture the difference in the distribution of distance values between proximities (already noted in Figure \ref{fig:dist_hist}). In particular, the MDS plots for Random Forest are essentially a disk, with a higher density of points on the boundary. This is due to the fact that Random Forest proximity tends to place most points the maximum distance (1 unit) apart from each other. By contrast, the MDS plot for QCML is more compact, with a center core consisting of lower-yield bonds and outliers corresponding to higher-yield bonds. 

MDS can also be used to visualize and compare the rankings of the top $k$ neighbors of an individual bond, for both QCML and Random Forest proximity. In Figure \ref{fig:bond_plots} (a), a reference bond from HYG is plotted, along with its top 10 neighbors for both QCML and Random Forest proximity. This bond has 50\% yield, making it an outlier that lies in a particularly sparse region of the data set. This sparsity is clear in the MDS plot of the QCML proximities (top of Figure \ref{fig:bond_plots} (a) ).  Nevertheless, among the top 10 nearest neighbors QCML identifies a similar bond also with 50\% yield. In contrast, most of the 10 neighbors for Random Forest are relatively far from the reference bond, resulting in higher yield prediction error (18.17\% prediction for Random Forest vs. 33.04\% for QCML). In Figure \ref{fig:bond_plots} (b), the same experiment is repeated for IGSB. In this case the reference bond chosen has very low spread, and is situated in the center core of the data. For this example, Random Forest is able to find 10 close neighbors and the spread approximation (11.2 basis points) is better than that of QCML (24.3 basis points). 

In general, one can expect QCML to outperform within a cohort of bonds that lie in sparse patches of the data, due to the ability of QCML to find a more compact representation. In contrast, Random Forest proximities might exacerbate sparsity by placing bonds as far from each other as possible. While these effects are not very noticeable in IG bonds, which tend to cluster around a center core of low yield bonds, they feature prominently in HY bonds, which are more scattered. This is very likely the source of advantage of QCML proximity that was noted earlier in the high-yield space.

\section{Conclusion}
\label{section:conclusion}

This study has shown that QCML, a novel paradigm for machine learning rooted in the mathematical formalism of quantum theory, on the whole performs comparably to, and in some cases better than, traditional machine learning methods in the context of both supervised regression and distance metric learning tasks. Although we do not make any claim that QCML is uniformly superior to traditional ML methods, our analysis has shown that there exist multiple contexts in which it can generate superior performance. Moreover, our results suggest that these methods can provide a real, practical advantage in the financial application of identifying trade-able alternatives to illiquid securities. This was most clearly illustrated in the context of identifying tradable alternatives for high yield corporate bonds in the HYG index, where the alternative bonds indicated by the QCML metric are substantially closer in yield to the desired bond than are those indicated by distance metrics based on random forests or Euclidean distance. We hope that this investigation motivates further study of the practical applications, both financial and otherwise, of this novel and exciting paradigm in machine learning.  

\section*{Acknowledgement}
The views expressed here are those of the authors alone and not of BlackRock, Inc. The contents of this article do not constitute investment advice.
%The example results shown in this paper are based on academic settings and hypothetical scenarios and are strictly for demonstration purposes. These results must not be taken seriously for any investment purposes whatsoever.

\bibliographystyle{plain}
\bibliography{refs}

\end{document}